\documentclass[10pt,reqno,oneside]{amsart}
\usepackage{verbatim,amsmath,amssymb,cite,epsfig,subfigure,color,hyperref}

\numberwithin{equation}{section}

\newcommand{\Real}{\mathbb R}

\newcommand{\supp}{\operatorname{supp}}
\title[Equilibrium Energy and Entropy of Vortex Filaments]
{Equilibrium Energy and Entropy of Vortex Filaments on a Cubic Lattice: A Localized Transformations Algorithm}

\author{Pavel B\v{e}l\'{\i}k}
\author{Eric Bibelnieks}
\author{Robert Laskowski}
\author{Aleksandr Lukanen}
\author{Douglas P.~Dokken}

\address{P.~B\v{e}l\'{\i}k\\
Department of Mathematics, Statistics \& Computer Science\\
Augsburg University\\
Minneapolis, MN\\
USA}
\email{belik@augsburg.edu}

\address{A.~Lukanen\\
Augsburg University\\
Minneapolis, MN\\
USA}

\address{R.~Laskowski\\
Augsburg University\\
Minneapolis, MN\\
USA}

\address{E.~Bibelnieks\\
Augsburg University\\
Minneapolis, MN\\
USA}

\address{D.~P.~Dokken\\
Department of Mathematics\\
University of St.~Thomas\\
St.~Paul, MN\\
USA}
\email{dpdokken@stthomas.edu}



\date{\today}

\begin{document}

%%%%%%%%%%%%
% Abstract %
%%%%%%%%%%%%
\begin{abstract}
In this work we propose a new algorithm for the computation of statistical equilibrium quantities on a cubic lattice when both an energy and a statistical temperature are involved. We demonstrate that the pivot algorithm used in situations such as protein folding works well for a small range of temperatures near the polymeric case, but it fails in other situations. The new algorithm, using localized transformations, seems to perform well for all possible temperature values. Having reliably approximated the values of equilibrium energy, we also propose an efficient way to compute equilibrium entropy for all temperature values. We apply the algorithms in the context of suction or supercritical vortices in a tornadic flow, which are approximated by vortex filaments on a cubic lattice. We confirm that supercritical (smooth, ``straight'') vortices have the highest energy and correspond to negative temperatures in this model. The lowest-energy configurations are folded up and ``balled up'' to a great extent. The results support A.~Chorin's findings that, in the context of supercritical vortices in a tornadic flow, when such high-energy vortices stretch, they need to fold.
\end{abstract}

%%%%%%%%
% Body %
%%%%%%%%
\allowdisplaybreaks
\thispagestyle{empty}
\maketitle

\section{Introduction}
\begin{figure}
    \includegraphics[width=0.7\textwidth]{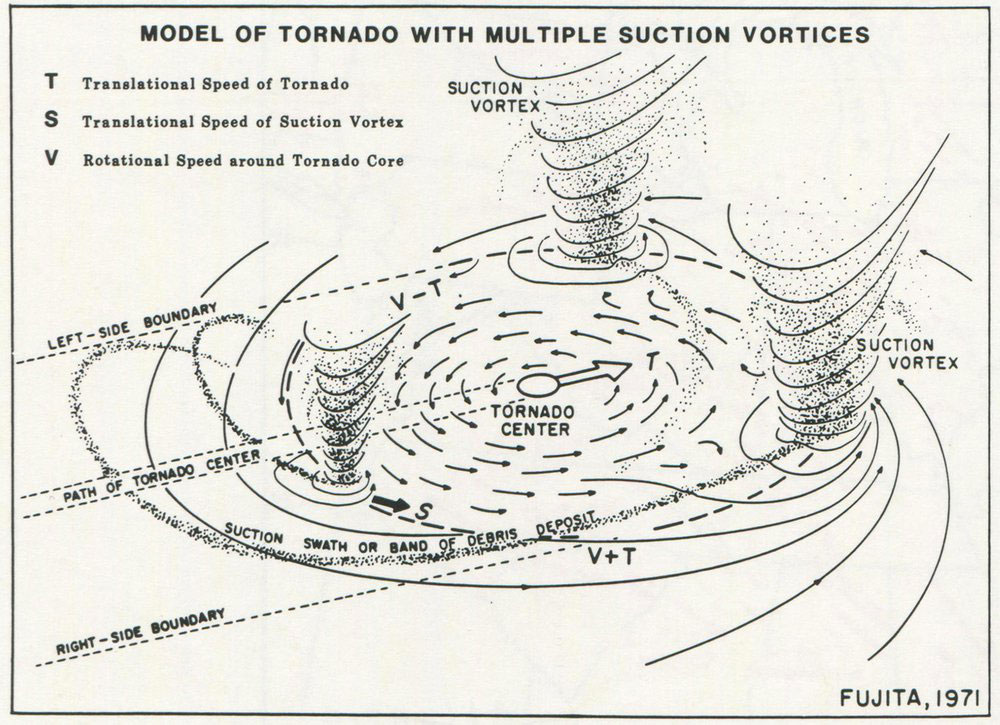}
    \caption{A visualization of a flow in a strong tornado with multiple strong suction vortices due to Tetsuya ``Ted'' Fujita \cite{fujitatribute}.}
    \label{fig:suctionvorticesFujita}
\end{figure}

\begin{figure}
    \includegraphics[width=0.9\textwidth]{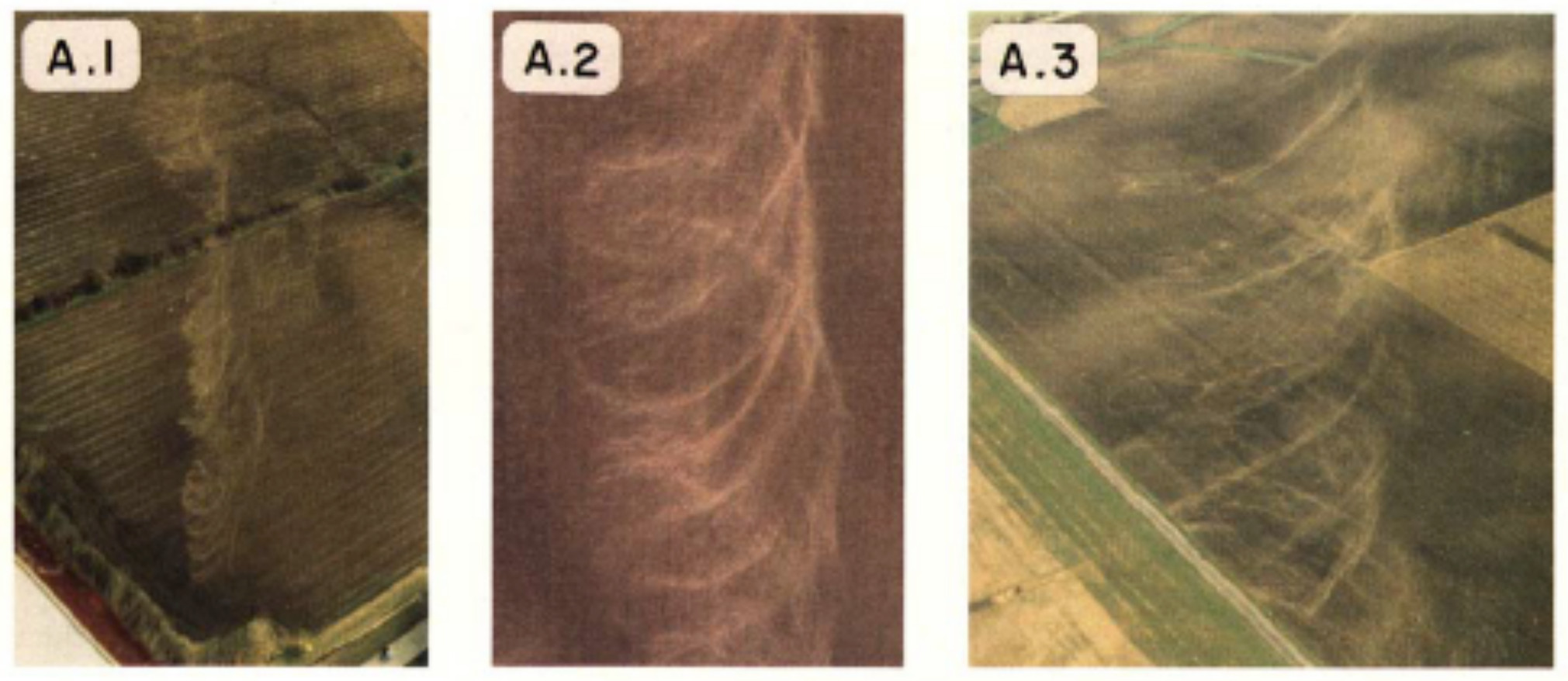}
    \caption{Tracks left in corn fields showing vortices spiraling into tornadoes and then dissipating. Locations and dates of occurrences are: (A.1) Decatur, Illinois tornado, April 3, 1974; (A.2) Magnet, Nebraska tornado, May 6, 1975; (A.3) Homer Lake, Indiana tornado, April 3, 1974. \copyright~American Meteorological Society, \cite{fujita81}.}
    \label{fig:suctionspots}
\end{figure}

During the formation stage of a tornado, long narrow vortices often appear spontaneously in the region of tornado formation, then fold up and dissipate. These vortices also appear in other stages of the tornado's existence. An example of such vortices in a strong tornado is shown schematically in Fig.~\ref{fig:suctionvorticesFujita} \cite{fujitatribute}. As indicated in the figure, such vortices are called {\it suction vortices}. Evidence for them can be seen in tornado videos or in the damage surveys done after the tornado has passed. Tracks of these vortices can be seen in damage surveys, where grass in lawns or plants in farmers' fields have been ripped from the ground, indicating their high energy density~\cite{fujita81} (see Fig.~\ref{fig:suctionspots}). Such strong, narrow vortices have been analyzed in \cite{fiedlerrotunno86} as so-called {\it supercritical vortices}. Their formation is related to the breakdown in the cyclostrophic balance, where the pressure-gradient force dominates, and the vortex collapses to a narrow filament; as the vortex collapses the energy density increases. This mirrors the behavior of negative-temperature vortices studied by A.~Chorin in his work on turbulence~\cite{chorin88,chorin90,chorin91,chorinakao91,chorin}. In these works, the negative-temperature vortices are straight and as they transfer energy to the surrounding flow, they fold up and dissipate. The similarity between the behavior of supercritical vortices studied in~\cite{fiedlerrotunno86} and negative-temperature vortices studied by Chorin is striking and is one of the motivations for this paper.

\begin{figure}
    \includegraphics[width=0.85\textwidth]{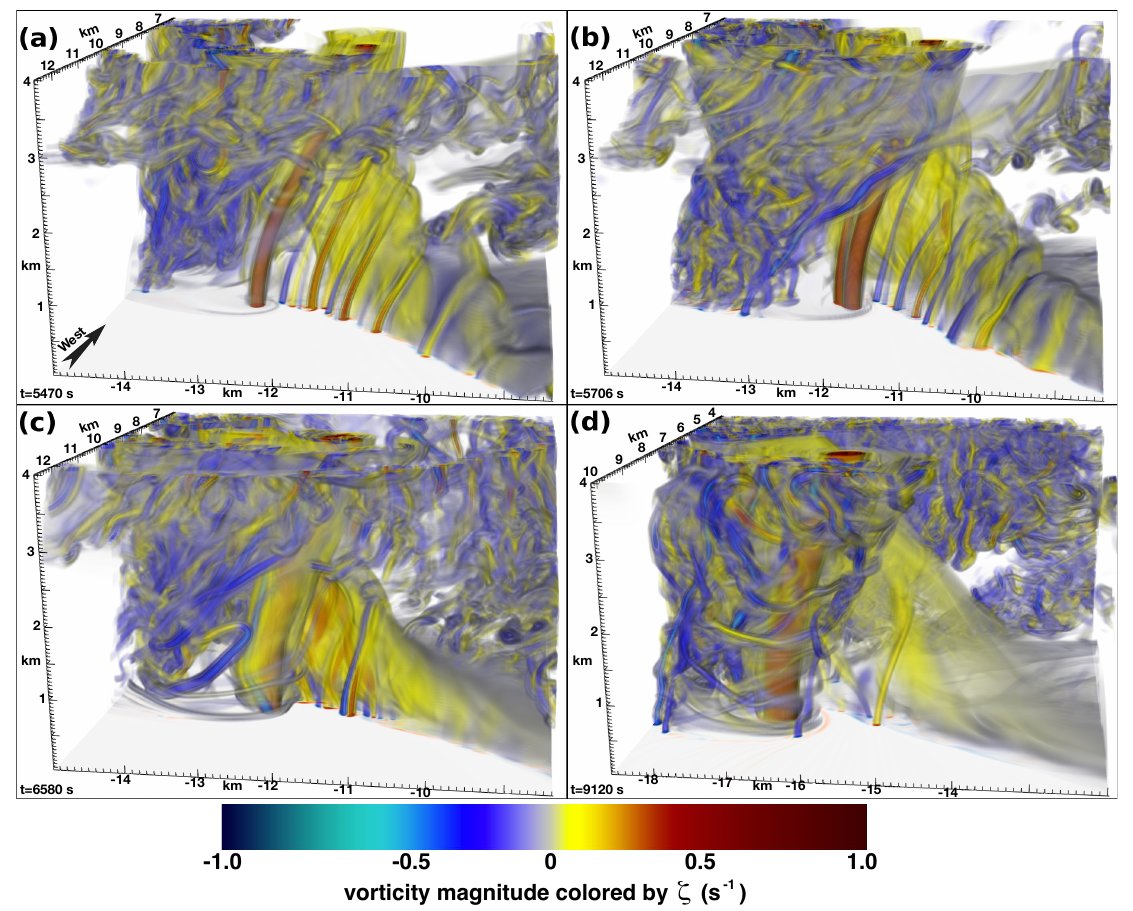}
    \caption{A computer-simulated time evolution of a tornadic storm in which vertical vorticity is provided to the tornado by intense, narrow, vertical vortices moving into it~\cite{orf17}. Those vortices then fold up, transfer their energies to the larger tornadic flow and dissipate. Physical times shown are (a) $t=5470$ s, (b) $t=5706$ s, (c) $t=6580$ s, and (d) $t=9120$ s. \copyright~American Meteorological Society, \cite{orf17}.}
    \label{fig:orf}
\end{figure}

The idea of supercritical or suction vortices playing an important role in tornadogenesis and tornado maintenance has also been discussed in~\cite{belikdokkenpotvinscholzshvartsman17} and demonstrated in a state-of-the-art numerical simulation~\cite{orf17}. A small snapshot of the simulated dynamical behavior of these vortices is shown in Fig.~\ref{fig:orf}. Notice how the intense, narrow, vertical vortices to the right of the developing or existing tornado move into the region where the tornado is forming or has formed, and how those vortices eventually fold and dissipate, transferring their energy to the surrounding flow. Another instance of smaller, violent vortices within a large tornadic flow is the May 31, 2013 El Reno, OK tornado analyzed and discussed in~\cite{bluesteinthiemsnyderhouser18,wurman_BAMS14}. The radar data, obtained by a Doppler on Wheels mobile radar, reveal the existence of several intense multiple vortices within the larger flow. A possible whole hierarchy of vortices within vortices is discussed in~\cite{belikdahldokkenpotvinscholzshvartsman18}.

Motivated by Chorin's results, our goal is to reliably compute the energy and entropy of vortex filaments on a cubic lattice and use this knowledge in the context of supercritical suction vortices and their behavior. To this extent, we employ the model developed in~\cite{chorin88,chorin90,chorin91,chorinakao91,chorin}, in which a vortex filament is modeled as a self-avoiding walk (SAW) on a cubic lattice and its energy is readily computed. This model was further extended to vortex structures with Brownian cores and fractal cross sections~\cite{flandoli02} and to Brownian semimartingales~\cite{flandoligubinelli02}. Another model rigorously studies an ensemble of nearly parallel vortices, however, under the restriction that the vortex filaments cannot fold~\cite{lionsmajda00}, a restriction that is detrimental to our consideration of vortices that fold up and dissipate. The cubic lattice SAW model used by Chorin has yielded results in a narrow range of statistical temperatures due to the employment of a Markov chain Monte Carlo (MCMC) algorithm (the {\it pivot algorithm}~\cite{lal69,madrassokal88}) that is well suited for polymers (infinite temperature or maximum entropy case), but fails to deliver reliable results when this is not the case. We propose a new algorithm, the {\it localized transformations} algorithm, that appears to alleviate most of the problems experienced by the pivot algorithm. The new algorithm allows us to compute equilibrium average energies that can be validated when exact values are known, and at the moment we do not have any indication that the results are significantly off in general. Having computed energies that appear reliable, we also propose a way to efficiently compute the entropy of the system whose accuracy is mainly affected by how accurately the average energies have been computed.

\section{Background Mathematics}
\label{sec:background}
Atmospheric flows can be modeled using Euler's equations relating velocity, pressure, and mass density of the fluid and external body forces. Typical flows are usually incompressible, so the divergence of the velocity field is zero. Isentropic flows are nearly incompressible when one of two scenarios occurs: either the flow speeds or the local changes in flow speeds along streamlines are small compared to the speed of sound in the medium~\cite{chorinmarsden93}. A numerical study of intense tornadic compressible and incompressible isentropic flows has shown little difference in results \cite{xialewellens03}.

With $\bf u$ being the fluid's velocity, $\rho$ its mass density, $p$ its pressure, and $\bf b$ an external body force, all functions of position $\bf{x}\in\Real^3$ and time $t\in\Real$, the governing equations for an incompressible fluid flow are
\begin{gather}
\label{eq:euler}
  \frac{D{\bf u}}{Dt}
  =
  -\frac{1}{\rho}\nabla p+{\bf b},\qquad
  \nabla\cdot{\bf u}=0,\qquad
  \dfrac{D\rho}{Dt}=0,
\end{gather}
where $\dfrac{Df}{Dt}=\dfrac{\partial f}{\partial t}+({\bf u}\cdot\nabla)f$ denotes the material derivative of a scalar function $f$. When applied to a vector function, the operator $\dfrac{D}{Dt}$ applies component wise.

The vorticity field $\boldsymbol\xi$ of the velocity field $\bf u$ is given by the curl of $\bf u$, i.e., ${\boldsymbol\xi}=\nabla\times{\bf u}$. By applying curl to the first equation in \eqref{eq:euler} one can obtain an equation for $\boldsymbol\xi$,
\begin{equation*}
  \label{eq:vorticity_eqn}
  \frac{\partial\boldsymbol\xi}{\partial t}
  =
  \nabla\times({\bf u}\times\boldsymbol\xi)
  +
  \frac{1}{\rho^2}\nabla\rho\times\nabla p
  +
  \nabla\times{\bf b}.
\end{equation*}
In this equation, the first term in the right-hand side corresponds to the ``barotropic'' generation of vorticity (capturing the advection, stretching, and tilting of the vertical vorticity; see, e.g., \cite{klemp87}), while the second term corresponds to the ``baroclinic'' generation of vorticity, i.e., vorticity generation due to the misalignment of the gradients of mass density and pressure. The body force is often conservative (e.g., due to gravity), in which case $\nabla\times{\bf b}={\bf 0}$.

In what follows, we will focus on flows in which vorticity is supported on long, narrow vortex tubes embedded in a larger irrotational flow. These may be baroclinic or barotropic in origin, and may demonstrate themselves, e.g., as the suction vortices discussed earlier~\cite{fujita81}.

Given a sufficiently fast decaying vorticity field ${\boldsymbol\xi}({\bf x})$ in $\mathbb{R}^3$, the system
\begin{equation*}
  {\boldsymbol\xi}
  =
  \nabla\times{\bf u},
  \qquad
  \nabla\cdot{\bf u}
  =
  0
\end{equation*}
can be solved for the velocity field \cite{majdabertozzi01}
\begin{equation*}
  {\bf u(x)}
  =
  -\frac{1}{4\pi}\int\frac{({\bf x-x'})\times{\boldsymbol\xi}({\bf x'})}{|{\bf x-x'}|^3}\,d\bf x'.
\end{equation*}
This expression is known as the Biot--Savart law. From this, one has an expression for the kinetic energy of the flow~\cite{lamb}
\begin{align*}
  E
  &=
  \frac{1}{2}\int_{\Real^3}\rho({\bf x})|{\bf u(x)}|^2\,d{\bf x}\\
  &=
  \frac{1}{8\pi}\iint_{\Real^3\times\Real^3}\rho({\bf x})\,\frac{{\boldsymbol\xi}({\bf x})\cdot{\boldsymbol\xi}({\bf x'})}{|{\bf x-x'}|}\,d{\bf x'}\,d{\bf x}
  +
  \frac{1}{8\pi}\iint_{\Real^3\times\Real^3}\frac{(\nabla\rho({\bf x})\times{\bf u(x)})\cdot{\boldsymbol\xi}({\bf x'})}{|{\bf x-x'}|}\,d{\bf x'}\,d{\bf x}.
\end{align*}
For a homogeneous fluid rescaled so that $\rho({\bf x})\equiv1$ the second term above vanishes and the expression for the kinetic energy becomes \cite{chorin}
\begin{equation}
\label{eq:energy_3D}
  E
  =
  \frac{1}{8\pi}\iint_{\Real^3\times\Real^3}\frac{{\boldsymbol\xi}({\bf x})\cdot{\boldsymbol\xi}({\bf x'})}{|{\bf x-x'}|}\,d{\bf x'}\,d{\bf x},
\end{equation}
where we note, however, that the integral can be taken over $\supp{\boldsymbol\xi}\times\supp{\boldsymbol\xi}$, so the integrals are evaluated over the vortex tubes only.

Following Chorin's work, we now define an approximation to the energy \eqref{eq:energy_3D} for an infinitely thin vortex filament, a ``vortex line.'' The assumption is that the vortex tube can be approximately divided into $N$ narrow circular cylinders $I_i$ of equal length, for which we then have
\begin{align*}
  E
  &=
  \frac{1}{8\pi}\iint_{\Real^3\times\Real^3}\frac{{\boldsymbol\xi}({\bf x})\cdot{\boldsymbol\xi}({\bf x'})}{|{\bf x-x'}|}\,d{\bf x'}\,d{\bf x}\\
  &=
  \frac{1}{8\pi}\sum_i\sum_{j\ne i}\int_{I_i}\int_{I_j}\frac{{\boldsymbol\xi}({\bf x})\cdot{\boldsymbol\xi}({\bf x'})}{|{\bf x-x'}|}\,d{\bf x'}\,d{\bf x}
  +
  \frac{1}{8\pi}\sum_i\int_{I_i}\int_{I_i}\frac{{\boldsymbol\xi}({\bf x})\cdot{\boldsymbol\xi}({\bf x'})}{|{\bf x-x'}|}\,d{\bf x'}\,d{\bf x},
\end{align*}
where the first term in the last expression corresponds to interactions of one of the cylinders with the others and is referred to as {\it interaction} or {\it exchange} energy, while the second term is referred to as {\it self-energy} and gives the contribution to the kinetic energy arising from interactions of nearby points along the vortex filament. In Chorin's work, the self-energy term is neglected. For discussions of this term see, e.g., \cite{chorin90,chorin91,chorinakao91,chorin}.

Finally, for an infinitely thin vortex line, the exchange energy is approximated in the following way. Assume that the vortex consists of $N$ linear segments $I_i$ of equal length, whose midpoints are denoted by ${\bf M}_i$. Since vorticity along a vortex line is parallel to it, let ${\boldsymbol\xi}_i$ denote the ``vorticity'' vector on the segment $I_i$, which is in the direction of $I_i$ and has magnitude equal to the length of the segment $I_i$. Then the exchange energy is approximated by
\begin{equation}
\label{eq:energy}
  E
  =
  \frac{1}{8\pi}\sum_{i}\sum_{j\ne i}\frac{{\boldsymbol\xi_i}\cdot{\boldsymbol\xi_j}}{|{\bf M}_i-{\bf M}_j|}.
\end{equation}
This scaling corresponds to the vortex having circulation equal to one \cite{chorin90,chorin91}. In the next section we will discuss the cubic lattice approximation, in which the individual segments $I_i$ connect two nearest neighbors on a cubic lattice, so only angles of $90^\circ$ and $180^\circ$ between neighboring segments will be allowed. Notice that due to the dot products in \eqref{eq:energy} the largest contribution to the energy comes from nearby segments oriented in the same direction, smallest (negative) contribution from nearby segments oriented in the opposite direction, and orthogonal segments contribute zero to the energy. That is, the largest energy will correspond to straight vortices, while smallest energy will likely correspond to very folded up vortices.

\section{Vortex Filaments on a Cubic Lattice}
\label{sec:lattice}
To simplify the set of all possible configurations of a line vortex in 3D, we only consider those constrained to a cubic lattice $\mathbb Z^3$ with lattice constant one~\cite{chorin88,chorin90,chorin91,chorinakao91,chorin}. The vortex filament will then correspond to a self-avoiding random walk (SAW) on this lattice, a concept of great interest in the polymer and protein community (see, e.g., \cite{bachmann14}). Even with this simplifying assumption, the problem of studying various vortex configurations is intractable exactly, since the number of all possible SAWs of $N$ segments appears to grow exponentially as a function of $N$ \cite{macdonaldjosephhuntermoseleyjanguttman00,schiemannbachmannjanke05,sbb11}. The exact enumeration of all possible configurations has been achieved for only small values of $N$ and the number has been increasing slowly with time ($N=26$ in 2000~\cite{macdonaldjosephhuntermoseleyjanguttman00} and $N=36$ in 2011 \cite{sbb11}). In Table~\ref{tab:SAWs} we show the number of SAWs of $N$ steps on a cubic lattice for $N=1,\dots,9$, $36$ (all exact), and $1000$ (an estimate based on a best-fit formula~\cite{sbb11}).
\begin{table}
  \begin{tabular}{c|c}
    $ N$ & $M$, the number of SAWs of length $N$\\
    \hline
    $ 1$ & $6$\\
    $ 2$ & $30$\\
    $ 3$ & $150$\\
    $ 4$ & $726$\\
    $ 5$ & $3\,534$\\
    $ 6$ & $16\,926$\\
    $ 7$ & $81\,390$\\
    $ 8$ & $387\,966$\\
    $ 9$ & $1\,853\,886$\\
    \dots& \dots\\
    $36$ & $2\,941\,370\,856\,334\,701\,726\,560\,670\approx2.9\times10^{24}$\\
    $1000$ & $\approx1.5\times10^{671}$\\
  \end{tabular}
  \vskip0.5\baselineskip
  \caption{Number of SAWs on length $N$ on a cubic lattice. The values for $N\le36$ are exact~\cite{sbb11} and the value for $N=1000$ is an estimate based on a best-fit curve~\cite{sbb11}.}
  \label{tab:SAWs}
\end{table}

In this paper we focus on the demonstration of an improvement of the computation of average energy and entropy compared to previous attempts, and thus fairly small values of $N$ ($\le1000$) will be considered. These are small values compared to the computational state of the art, however the numbers of distinct SAWs, $M$, are still large, ranging from $M\approx10^{67}$ for $N=100$ to $M\approx10^{671}$ for $N=1000$ \cite{schiemannbachmannjanke05,sbb11}. As such, a statistical approach to our problem is required and an appropriate Monte Carlo technique with a suitable sampling procedure is usually employed.

In the next three sections we describe the relevant statistical mechanics, the pivot algorithm used in many simulations of SAWs, and the algorithm's modification needed to obtain accurate results.

\section{Statistical Mechanics Background}
\label{sec:statmech}
We now briefly review relevant concepts from statistical mechanics. Given a positive integer $N$ corresponding to the number of segments in a self-avoiding walk on a cubic lattice, we denote by $M$ the total number of possible SAWs starting at the origin. We denote by $x_i$, $i=1,\dots,M$, the possible SAWs, or vortex configurations of length $N$. The vortex configurations $x_i$ represent individual states in the system $\mathcal{S}_N=\{x_i:\ i=1,\dots, M\}$.

The energy of each configuration, a state energy, will be denoted by $E_i=E(x_i)$ with $E$ defined in \eqref{eq:energy}. The probability $p_i$ that a system will be in state $x_i$ is given by the Boltzmann probability distribution
\begin{equation}
\label{eq:boltzmann_distribution}
  p_i
  =
  \frac{e^{-\beta E_i}}{Z},
  \qquad
  Z
  =
  \sum_{i=1}^Me^{-\beta E_i},
\end{equation}
where $Z$ is called the (canonical) partition function and $\beta=\dfrac{1}{k_BT}$ is sometimes referred to as ``coldness'' as it is proportional to the reciprocal of the (statistical) temperature, $T$. The Boltzmann constant, $k_B$, will be assumed equal to $1$ in subsequent computations but we will use it in the general results in this section. Note that when $\beta=0$ all states are equally likely and $p_i=1/M$ for $i=1,\dots,M$. This is sometimes referred to as the {\it polymeric} case as equally likely configurations are considered when modeling the behavior of polymers~\cite{bachmann14}.

Notice that the relationship $\beta=\dfrac{1}{k_BT}$ can be also used to {\it define} the temperature $T=\dfrac{1}{k_B\beta}$. Consider $\beta$ increasing from $-\infty$ to $+\infty$. When $\beta$ increases from $-\infty$ to $0$, this corresponds to $T$ decreasing from $0$ to $-\infty$; when $\beta$ increases from $0$ to $+\infty$, this corresponds to $T$ decreasing from $+\infty$ to $0$. Put together, when one considers the temperature $T$, temperature first increases from $T=0$ to $T=+\infty$; considering the one-sided limits as $\beta\to0$, one can identify $T=+\infty$ with $T=-\infty$ and simply write $T=\infty$. Then temperature further increases as $T$ changes from $T=\infty$ through negative values towards $0$. Thus negative temperatures are higher than positive temperatures in this sense. (This idea corresponds to traversing a circle obtained by transforming the real number line into a circle by identifying $+\infty$ and $-\infty$, both for $\beta$ and $T$; this identification is a special case of the M\"{o}bius transformation of the complex plane.) Another explanation for why negative temperatures are higher than positive ones is based on energy (and entropy) considerations. For further details, see \cite{belikdokkenpotvinscholzshvartsman17,landaulifshitz}.

Using the Boltzmann distribution \eqref{eq:boltzmann_distribution}, we can now define in a standard way the average of any function of the states $x_i\in\mathcal{S}_N$ for any finite $\beta$. For example, the average energy $\langle E\rangle$ of the system $\mathcal{S}_N$ at a given $\beta$ (or temperature $T$) is given by
\begin{equation}
\label{eq:average_energy}
  \langle E\rangle
  =
  \sum_{i=1}^Mp_iE_i.
\end{equation}
One of the goals of this paper is to reliably approximate $\langle E\rangle$ for any finite $\beta$, which would extend the results of Chorin (see, e.g., \cite{chorin91,chorinakao91,chorin}) that appear reliable only for a small interval of values of $\beta$ containing zero.

The Gibbs entropy of the system $\mathcal{S}_N$ at a given $\beta$ is given by
\begin{equation}
\label{eq:entropy}
  S
  =
  -k_B\sum_{i=1}^Mp_i\log p_i,
\end{equation}
which can also be viewed as an average quantity, since one can write
\begin{equation*}
  S
  =
  \sum_{i=1}^Mp_i\left(-k_B\log p_i\right)
  =
  \sum_{i=1}^Mp_iS_i
\end{equation*}
and view $S_i=-k_B\log p_i$ as an ``entropy'' of the state $x_i$. Also note that if $\beta=0$, then, since $p_i=1/M$, we have $S=k_B\log{M}$, which corresponds to Boltzmann's definition of entropy.

Taking a partial derivative of the logarithm of the partition function in \eqref{eq:boltzmann_distribution} with respect to $\beta$, we obtain
\begin{equation}
\label{eq:partial_of_Z}
  -\frac{\partial\log{Z}}{\partial\beta}
  =
  \frac{1}{Z}\sum_{i=1}^M E_i e^{-\beta E_i}
  =
  \langle E\rangle.
\end{equation}
Using \eqref{eq:boltzmann_distribution}, expression \eqref{eq:entropy} can be written as
\begin{equation*}
  S
  =
  \sum_{i=1}^M p_i(k_B\beta E_i+k_B\log{Z})
  =
  k_B\beta\langle E\rangle
  +
  k_B\log{Z},
  \label{eq:entropy2}
\end{equation*}
and differentiating it with respect to $\langle E\rangle$ and using \eqref{eq:partial_of_Z} gives the well-known result
\begin{equation}
  \frac{\partial S}{\partial\langle E\rangle}
  =
  k_B\frac{\partial\beta}{\partial\langle E\rangle}\langle E\rangle
  +
  k_B\beta
  +
  k_B\frac{\partial\log{Z}}{\partial\beta}\frac{\partial\beta}{\partial\langle E\rangle}
  =
  k_B\beta
  =
  \frac{1}{T}.
  \label{eq:dSdE}
\end{equation}
Finally, the Helmholtz free energy of the system is
\begin{equation*}
  F
  =
  \langle E\rangle
  -
  TS
  =
  \sum_{i=1}^Mp_i\left(E_i+k_BT\log p_i\right),
\end{equation*}
where, in view of \eqref{eq:boltzmann_distribution} and the relationship between $\beta$ and $T$, the term in parentheses equals $-k_BT\log{Z}$, i.e., a constant independent of $i$, and hence the expression for the Helmholtz free energy gives
\begin{equation}
  F
  =
  \langle E\rangle
  -
  TS
  =
  E_i+k_BT\log p_i
  \quad\text{ for any }i=1,\dots,M.
  \label{eq:helmholtz}
\end{equation}
In particular, this means that the entropy (and also the Helmholtz free energy and the partition function) can be computed from $\beta$, $\langle E\rangle$, $E_i$, and $p_i$ for any $i$ provided these values are known since \eqref{eq:helmholtz} can be rewritten as
\begin{equation}
\label{eq:entropy3}
  S
  =
  k_B\left(\beta\langle E\rangle-\beta E_i-\log{p_i}\right)
  \quad\text{ for any }i=1,\dots,M.
\end{equation}
Equations \eqref{eq:dSdE} and \eqref{eq:entropy3} will be used later in section \ref{sec:numerics}, where we discuss approximating the entropy of the vortex filament system.

\section{Monte Carlo Approach: The Pivot Algorithm}
\label{sec:pivot}
In order to reliably compute various statistical quantities of interest, such as the average energy \eqref{eq:average_energy} or the entropy \eqref{eq:entropy}, one usually employs Markov chain Monte Carlo (MCMC) techniques, since the space $\mathcal{S}_N$ of possible configurations of the vortex filament of length $N$ is too large for direct enumerations. As mentioned in section \ref{sec:lattice}, complete enumerations have been achieved only for small values of $N$ up to date \cite{macdonaldjosephhuntermoseleyjanguttman00,sbb11}.

An efficient and popular algorithm to generate a sequence of ``effectively independent'' SAWs is the {\it pivot algorithm} proposed by Lal in 1969 \cite{lal69} and analyzed and popularized by Madras and Sokal in 1988 \cite{madrassokal88}. In this algorithm, given an $N$-step SAW, one randomly (usually uniformly) picks a point on it as a ``pivot'' and applies a transformation from the symmetry group of the cubic lattice to the points subsequent to the pivot, using the pivot point as the origin. The points before the pivot stay intact. If the resulting walk is self-avoiding, it is accepted as the new element of the sequence; otherwise it is rejected and the original walk is repeated in the sequence. The symmetry group of the cubic lattice is the octahedral group $O_h$ with $48$ elements including the identity. The algorithm is ergodic and satisfies the condition of detailed balance \cite{madrassokal88}. This algorithm was used in the works that motivated our paper \cite{chorin90,chorin91,chorinakao91}, where only a small range of values of $\beta$ near $0$ (or large in absolute value temperatures $T$) was explored.

It is possible to use proper subgroups of $O_h$ that ensure ergodicity. For example, one can use a subset that contains all $90^\circ$ rotations and all axis reflections resulting in a set of $9$ transformations (if the identity is left out) \cite[Theorem 1]{madrassokal88}. In our limited comparison, the results using either $47$ transformations ($O_h$ without the identity) or the aforementioned $9$ transformations appeared similar when approximating average energy~\eqref{eq:average_energy}, so using fewer transformations might be preferable from efficiency point of view. We will refer to this algorithm as the P9 algorithm in what follows.

Various modifications of the pivot algorithm have been considered in the literature for various reasons. For example, additional possible moves useful in modeling polymers are discussed in \cite{bachmann14} and include ``end flips,'' ``corner flips,'' and ``crankshaft moves.'' Other modifications to speed up the algorithm have been described, for example, in \cite{kennedy08,clisby13}. Algorithms for random walks with fixed endpoints have also been proposed~\cite{madrasorlitskyshepp89}.

In the context of our problem we have discovered that the pivot algorithm struggles in many scenarios due to the involvement of the vortex filament energy~\eqref{eq:energy} in~\eqref{eq:boltzmann_distribution} and \eqref{eq:average_energy} and due to the fact that the pivot transformations are not local -- in any given step all of the points after the pivot are expected to change position leading to a possibly large change in the filament energy. As commented in~\cite{chorin91}, ``{\it For values of $|T|$ between $0.4$ and $1$ [\dots], the results are unreliable but suggestive, and for $|T|<0.4$ they are completely unreliable.}'' This leaves only a small range of $\beta$ values close to $0$ where the authors in~\cite{chorin91} feel confident in their results and leaves much room for improvement. We have made similar observations in our simulations using the pivot algorithm which led us to abandon the standard version of the algorithm and consider modifications and improvements to compute the average energy~\eqref{eq:average_energy} more accurately.

Specifically, we have observed that non-straight configurations at large (in absolute value) negative values of $\beta$ (for which the Boltzmann probability distribution~\eqref{eq:boltzmann_distribution} strongly favors high-energy, straight configurations) are extremely unlikely to straighten out in the MCMC simulation since intermediate steps are required that greatly affect the energy. Similarly, when starting with a straight configuration but $\beta$ is not large negative, the same need for an intermediate transformation with a large energy change leads to incorrect results.

We can quantify this observation with the following argument. Consider a filament of $N=2n$ segments such that all but the $(n+1)$st segment are in the same direction (the same argument applies to any other segment being different from the others). The energies of such filaments of various lengths $N=10,\dots,800$ computed using \eqref{eq:energy} are shown in the second row of Table~\ref{tab:energies}, denoted by $E$.
\begin{table}
  \begin{tabular}{c|*{8}{c}}
    %\hline
    $N$                  & $10$      & $20$      & $40$      & $80$      & $100$     & $200$     & $400$     & $800$     \\
    \hline
    $E$                  & $1.26965$ & $3.80838$ & $10.0521$ & $24.8048$ & $32.8638$ & $77.1223$ & $176.727$ & $398.055$ \\
    $E_\text{max}(N)$    & $1.53502$ & $4.13443$ & $10.4359$ & $25.2450$ & $33.3221$ & $77.6363$ & $177.296$ & $398.680$ \\
    $2E_\text{max}(N/2)$ & $1.02124$ & $3.07005$ & $8.26886$ & $20.8719$ & $27.8458$ & $66.6442$ & $155.273$ & $354.593$ \\
    $e^{\Delta E}$      & $0.78$    & $0.48$    & $0.17$    & $0.020$   & $7\times10^{-3}$ & $3\times10^{-5}$ & $5\times10^{-10}$ & $1\times10^{-19}$ \\
    $e^{10\Delta E}$    & $0.083$   & $6\times10^{-4}$ & $2\times10^{-8}$ & $8\times10^{-18}$ & $2\times10^{-22}$ & $3\times10^{-46}$ & $7\times10^{-94}$ & $2\times10^{-189}$ \\
    %\hline
  \end{tabular}
  \vskip0.5\baselineskip
  \caption{Values relevant for trying to turn an almost straight filament of length $N$ into a perfectly straight one. $E$ is the energy of a straight filament with one segment in the middle turned $90$ degrees; $E_\text{max}(N)$ is the energy of a straight filament; $2E_\text{max}(N/2)$ is the energy of a piecewise straight filament with a $90$-degree turn in the middle; the last two rows display the acceptance probabilities for $\beta=-1$ and $\beta=-10$ of a filament going from energy $E$ to energy $2E_\text{max}(N/2)$.}
  \label{tab:energies}
\end{table}
The third row of the table shows the energies of perfectly straight filaments, denoted by $E_\text{max}(N)$ (see \eqref{eq:max_energy} below). For large negative $\beta$ the straight configurations have larger probability \eqref{eq:boltzmann_distribution} of occuring, so the transition to the straight configuration should take place. However, given the set of possible transformations in the pivot algorithm, there is no way to transition directly to the straight configuration. At minimum, two transformations have to take place to turn the two $90$-degree angles along the $(n+1)$st segment to $180$-degree angles. The energy of one of the two possible intermediate configurations ($n=N/2$ steps in one direction followed by $n$ steps in an orthogonal direction) is shown in the fourth row of Table~\ref{tab:energies}, denoted by $2E_\text{max}(N/2)$. Not surprisingly, it shows a significant drop in energy compared to the two straight(er) configurations. To appreciate the energy difference between the starting and the intermediate configurations, in the last two rows of Table~\ref{tab:energies} we show the MCMC acceptance probabilities of the proposed transitions for $\beta=-1$ and $\beta=-10$. We note that even for a modest value of $N=200$ and $\beta=-1$, such a transformation is not very likely to be proposed and accepted, and it is highly unlikely that such a transformation would be proposed and accepted in a reasonable number of iterations even once for larger values of $N$. For $\beta=-10$ such a situation could arise as early as $N=40$.

At the other end of the $\beta$ spectrum, when the values of $\beta$ are large positive, the Boltzmann probability distribution strongly favors lowest-energy configurations, likely very much folded up and ``balled up'' or ``compressed'' in volume. We elaborate on this issue in section~\ref{sec:min_energy}. While the maximum-energy configurations are obviously the straight configurations with their energies easily computed for any $N$, the minimum-energy configurations are not known to us, nor are the actual minimum energies. However, based on the results shown in section~\ref{sec:pivot_results}, the MCMC simulation based on the pivot algorithm struggles to get anywhere close to the projected minimum values, thus begging for a different approach to the problem. Part of the problem with the pivot algorithm again is the need for intermediate configurations with unfavorable values of the energy; another problem, still possibly present in our approach, is that the minimum-energy configurations appear to require a particular structure that may be hard to get to from another ``balled up'' configuration using simple local transformations.

\section{Monte Carlo Approach: The Localized Transformations (LT) Algorithm}
\label{sec:localized}
The difficulties with the pivot algorithm outlined in the previous section, when energy and the Boltzmann probability distribution are involved, suggest that the pivot algorithm is too rigid in the non-polymeric cases ($\beta\ne0$ or $T\ne\infty$) and that different, more localized, transformations should be considered in order to better approximate average energies when $|\beta|>0$.

To this end, we propose and utilize an algorithm that we will refer to as the {\it localized transformations} (LT) algorithm that will use the following two types of transformations: one designed specifically to help with straightening out non-straight configurations at negative values of $\beta$, and one designed to help with compressing configurations in order to lower their energy at positive values of $\beta$. We next describe these two types of transformations.

As mentioned in the previous section, for large (in absolute value) negative values of $\beta$, the pivot algorithm is very unlikely to straighten out a configuration with all but one segment in one direction (or similar, more complicated configurations), since in order to remove the ``kink,'' the whole part after the kink has to be temporarily transformed as well, resulting in a significant change in energy. An easy solution would be to simply modify the one non-aligned segment while keeping the directions of the other segments the same. This still modifies the positions of the points after the kink, but it results in a small change in the filament energy. We generalize this idea to the following algorithm for an $N$-step SAW:
\begin{verbatim}
First Reconstruction Algorithm:
  1. Randomly select a number n_steps between 1 and N of consecutive steps to be modified.
  2. Randomly select a subwalk of length n_steps to be modified.
  3. Randomly reconstruct the selected subwalk.
  4. Connect the reconstructed subwalk to the remaining one or two pieces by translation.
\end{verbatim}
This algorithm is clearly ergodic since the reconstruction can be done on the whole SAW. Inverses of transformations are included and selected with equal probabilities, so the algorithm satisfies the detailed balance condition as well when the usual Monte Carlo acceptance probability is used. Practical considerations here include the choice of probability distributions in steps 1.~and 2.~(we used an exponential distribution in 1.~to emphasize shorter reconstructions and a uniform distribution in step 2.) and how much care should go into step 3.~(e.g., random reconstruction vs.~constructing a true SAW; we used a random reconstruction). It is easy to see that the configurations with one misaligned segment are easy to transform into a straight configuration with little energy change.

For large positive values of $\beta$, the filament configurations are expected to have small energy and be very folded up and ``balled up'' in order to lower their energy. The first reconstruction algorithm is not an ideal candidate here due to the fact that the part of the filament after the reconstructed subwalk is likely to be translated during the reconnection, likely leading to self-intersections of already balled up configurations. Motivated by transformations proposed for configurations with fixed endpoints~\cite{madrasorlitskyshepp89}, we propose a second reconstruction algorithm that reconstructs a subwalk while keeping its endpoints fixed:
\begin{verbatim}
Second Reconstruction Algorithm:
  0. Choose a number K between 2 and N for maximum number of allowed reconstructed steps.
  1. Randomly select a number n_steps between 2 and K of consecutive steps to be modified.
  2. Randomly select a subwalk of length n_steps to be modified.
  3. Permute the selected steps to generate a new subwalk.
  4. Connect the reconstructed subwalk to the remaining one or two pieces.
\end{verbatim}
Combining the two reconstruction algorithms results in an algorithm that is ergodic due to the ergodicity of the first algorithm. The second algorithm also clearly satisfies the detailed balance condition. Practical considerations here include the following. The number $K$ can be chosen arbitrarily, but we found that $K\approx N^{2/3}$ not only roughly corresponds to the number of surface points in a ``balled up'' configuration, it also appears to lead to reasonably small minimum energy values. For the probability distributions in 1.~and 2., we again used an exponential and a uniform distribution, respectively. The permutation in 3.~could be a random permutation; we found that choosing a (uniformly) random cyclic permutation led to better results, presumably due to a lower rejection rate. In general, our experience described below appears to lead to minimum computed energy values that are within fewer than $5$ percent of the predicted minimum values as shown in section~\ref{sec:min_energy}, which is a significant improvement over the pivot algorithm.

\section{Numerical Results}
\label{sec:numerics}
In this section we present our numerical results for the computation of the average energy $\langle E\rangle$ and the entropy $S$ as a function of the length of the vortex filament, $N$, and the inverse temperature, $\beta$. The motivation comes from extending the results presented in \cite{chorin91,chorinakao91,chorin} (and the references within) to cover a much broader range of values of $\beta$ than done previously. One of the goals is to demonstrate the improved performance of the new LT algorithm, not pushing the results to the largest possible lengths of the filaments. Thus our results extend only up to $N=1000$, but the range of values of $\beta$ for which we believe our results are reliable is significantly extended compared to \cite{chorin91,chorinakao91,chorin}.

In the individual subsections we present and discuss results for the average energy~\eqref{eq:average_energy} computed using the two algorithms, we also discuss and compute the entropy of the system, and we discuss the minimum energy configurations and how well they might be approximated by the new algorithm with the localized transformations.

\subsection{Energy Computations Using the Localized Transformations Algorithm}
\label{sec:localized_results}
In this section we present the numerical results for computing the average energy, $\langle E\rangle$, given in \eqref{eq:average_energy}, using the proposed algorithm with localized transformations that comprises of the first and second reconstruction algorithms described in the previous section.

The implementation details are as follows. As in \cite{chorin91}, our Monte Carlo simulations are started from straight filaments of $N$ steps, which, for a given $N$, are also the largest-energy configurations with energy
\begin{equation}
  E_{\text{max}}(N)
  =
  \frac{1}{4\pi}
  \left[
    N\sum_{k=1}^{N-1}\frac{1}{k}-(N-1)
  \right].
  \label{eq:max_energy}
\end{equation}
In light of \eqref{eq:boltzmann_distribution}, the straight configurations are the most likely configurations for negative temperatures ($\beta<0$), and we expect $\langle E\rangle\approx E_{\text{max}}$ when $\beta$ is large (in absolute value) negative. Therefore, we start our computation at one such $\beta$ ($\beta_\text{first}=-20$ for the results below) and compute an approximation of the average energy for this value of $\beta$ using the straight configuration as a starting point. We then increase $\beta$ by a small increment ($\Delta\beta=0.5$ in the results below), use the lowest-energy configuration encountered so far as a starting point, and compute $\langle E\rangle$ with the new $\beta$. We continue this way until a stopping value of $\beta$ has been achieved ($\beta_\text{last}=100$ in the results below).

The average energy for each value of $\beta$ is computed as follows. First, we allow for $100N$ ``burn-in'' transformations to forget recent history and then we perform $\max\{500N,\,200000\}$ averaging transformations to compute $\langle E\rangle$. Conceivably, these numbers could easily be made larger and potentially slightly affect the computed values.

During each averaging iteration, either the first or the second reconstruction algorithm is selected randomly with probability $1/2$, which could conceivably be modified based on the value of $\beta$. For the first reconstruction algorithm a ratio $r_1\in(0,1]$ is passed in and a geometric probability distribution with the ratio $r_1$ is constructed for the numbers $1$ through $N$. (If $r_1=1$, the distribution is uniform.) A random number {\tt n\_steps} is drawn from the set $\{1,\dots,N\}$ according to this probability distribution. Clearly, for $r_1<1$ it is more likely to reconstruct shorter subwalks than longer ones. Finally, a starting point of the subwalk to be reconstructed is chosen randomly uniformly from possible candidates. For the second reconstruction algorithm, a similar process is followed. With a ratio $r_2\in(0,1]$, first a geometric probability distribution with the ratio $r_2$ is constructed for the numbers $2$ through $K$, where $K$ is the integer nearest to $N^{2/3}$. Then a random number {\tt n\_steps} is drawn from the set $\{2,\dots,K\}$ according to this probability distribution. Finally, a starting point of the subwalk to be reconstructed is chosen randomly uniformly from possible candidates. At this point, a reconstruction of the subwalk is performed according to steps 3.-4.~in the relevant reconstruction algorithm and the new filament is tested for self-avoidance and for acceptance in the MCMC algorithm. If it is self-avoiding and accepted using the Boltzmann probability criterion, the newly constructed filament becomes the new filament in the Markov chain; otherwise the current one is repeated.

\begin{figure}
    \includegraphics[width=0.75\textwidth]{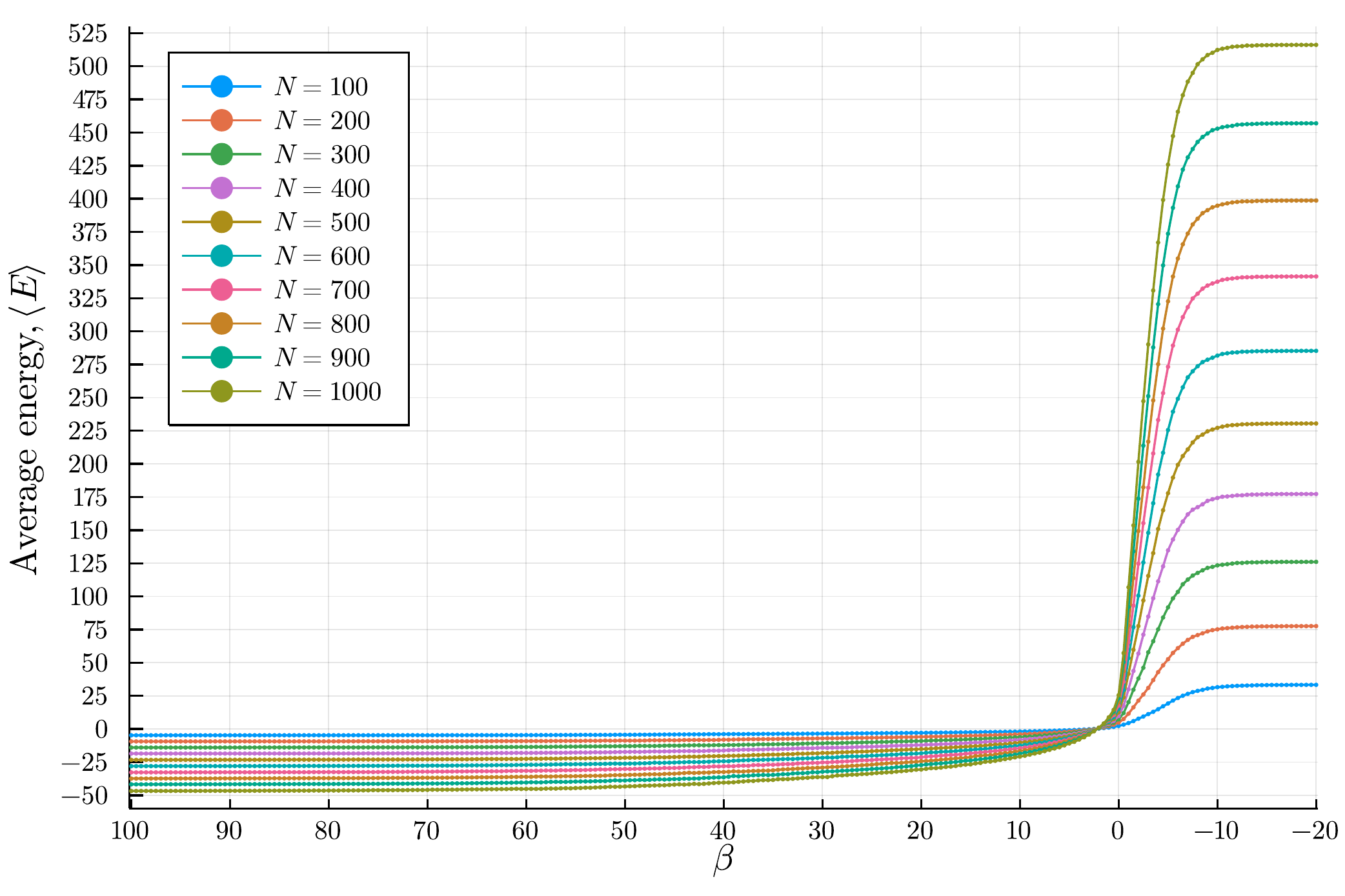}
    \caption{Average energies \eqref{eq:average_energy} for $N=100,\dots,1000$ computed using the LT algorithm.}
    \label{fig:results-100-1000}
\end{figure}

\begin{figure}
    \includegraphics[width=0.75\textwidth]{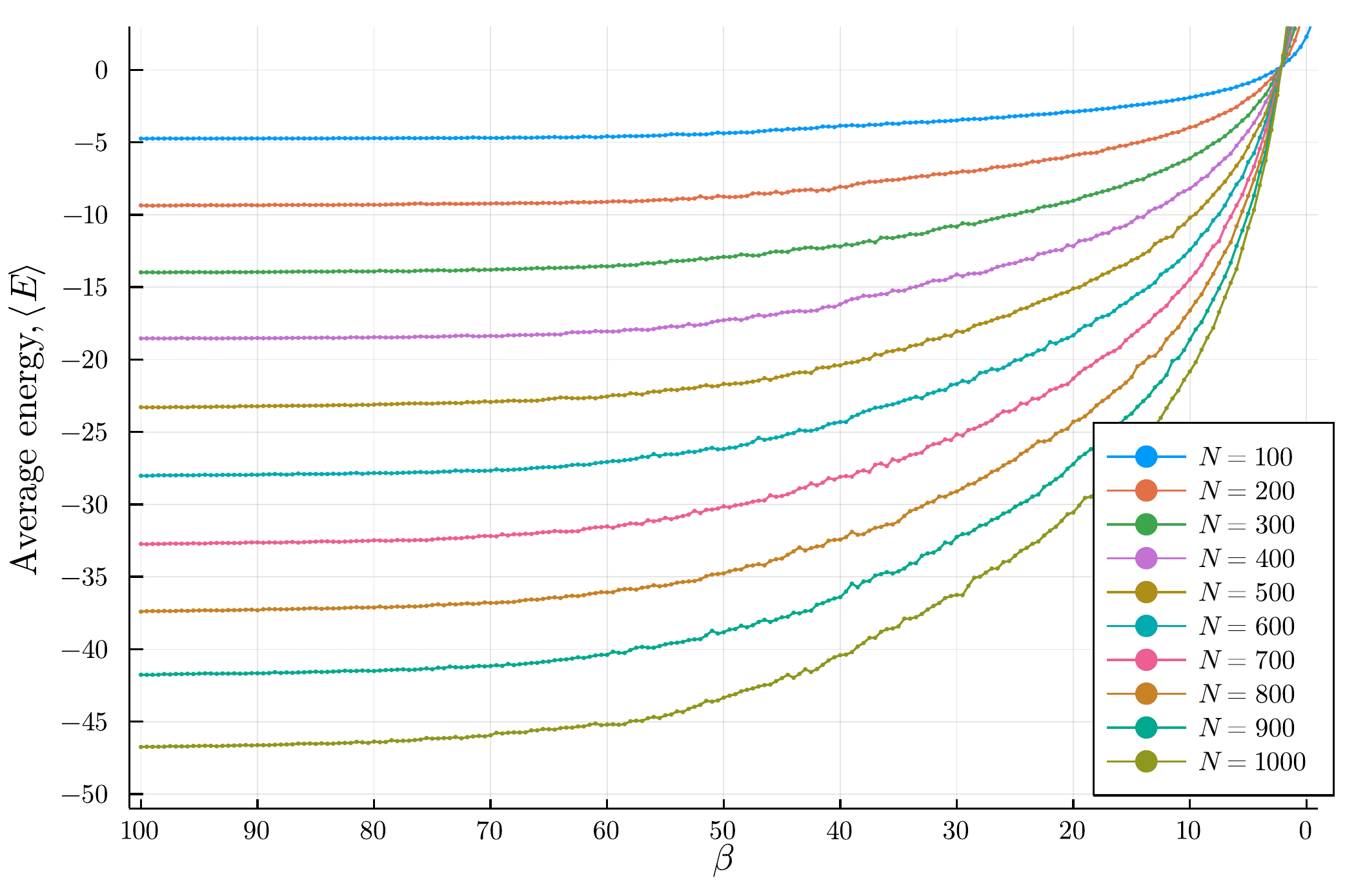}
    \caption{Same results as in Fig.~\ref{fig:results-100-1000} but displayed only for $\beta>0$.}
    \label{fig:results-100-1000_left-side}
\end{figure}

In Fig.~\ref{fig:results-100-1000} we show the computed results with $N=100$ through $N=1000$. We note that in this and subsequent figures $\beta$ intentionally runs from positive values on the left to negative values on the right so that temperature, $T=1/(k_B\beta)$, increases from left to right. For each value of $N$, the computation was repeated six times and averaged results are graphed. For each $N$, the average energy increases with temperature and, as expected, it levels off at its maximum value when $\beta<0$ and large in absolute value. Specifically, for $\beta\in[-20,-10)$ the average energies are near their maximum values \eqref{eq:max_energy}, corresponding to, on average, straight configurations. As temperature is lowered in the computation, average energy decreases as expected in light of \eqref{eq:boltzmann_distribution}. Due to the finite lattice spacing, for each $N$ there is a minimum energy a filament of $N$ steps can achieve, so it is not surprising that the average energies level off as $\beta\to\infty$. This is demonstrated in Fig.~\ref{fig:results-100-1000} and its zoomed-in version in Fig.~\ref{fig:results-100-1000_left-side}. We do not have an explicit expression for the minimum energy similar to \eqref{eq:max_energy}, but we provide some insights in section~\ref{sec:min_energy}. Notice that from \eqref{eq:max_energy} the maximum energies grow like $N\log{N}$, which is illustrated in Fig.~\ref{fig:results-100-1000_left-side}, but the computed results suggest a linear decay in the minimum energies.

\subsection{Energy Computations Using the Reduced Pivot (P9) Algorithm}
\label{sec:pivot_results}
As pointed out in section~\ref{sec:pivot}, the pivot algorithm is not suitable for our problem. In this section we provide some additional numerical evidence for this claim. We will attempt to reproduce some of the results from \cite{chorin91,chorinakao91,chorin}, discuss them, and provide a comparison to the results shown in the previous section generated using the new algorithm described in section~\ref{sec:localized}. Since not all implementation details are provided in~\cite{chorin90,chorin91,chorinakao91}, our comparison may be affected by our implementation choices, which are described below in detail.

\begin{figure}
    \begin{center}
        \includegraphics[width=0.7\textwidth]{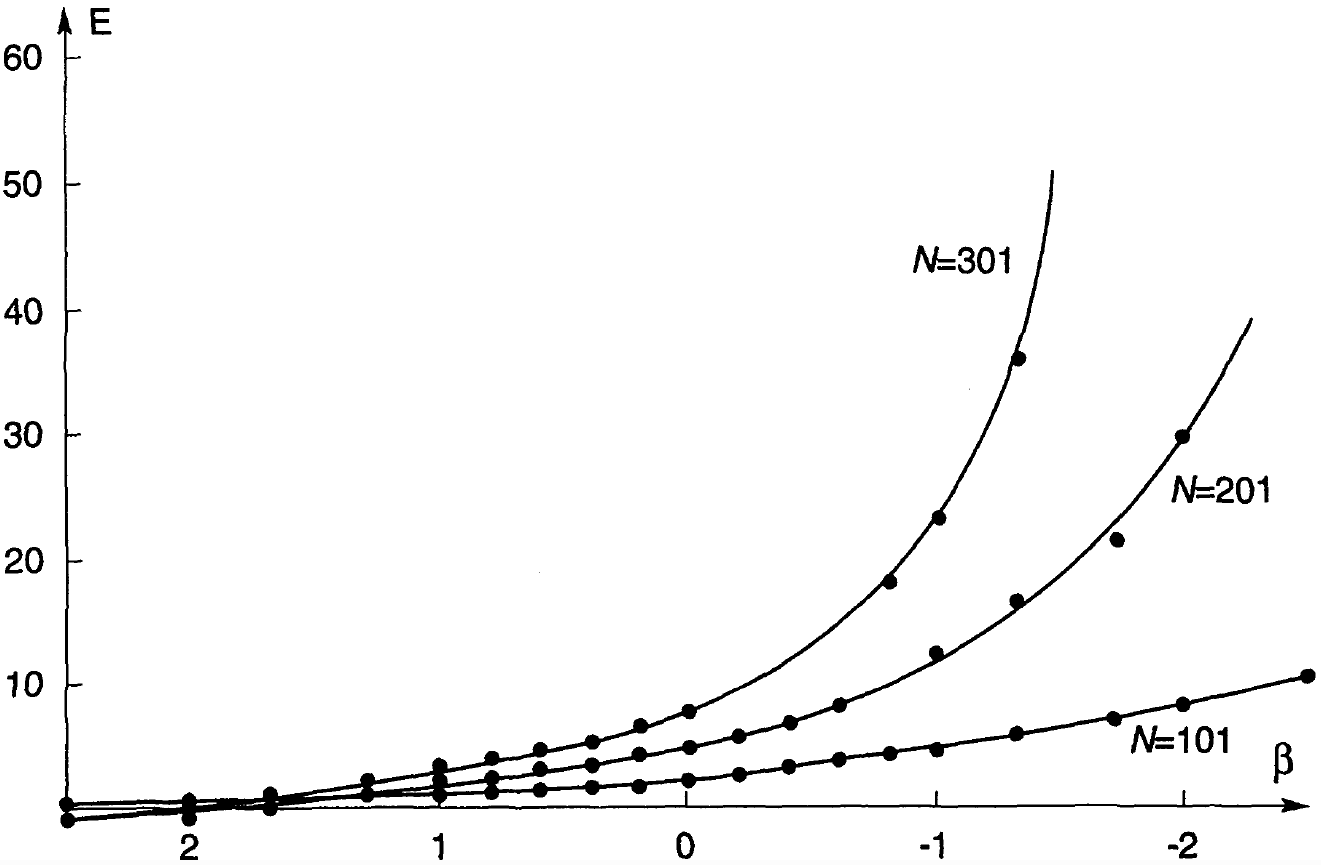}
    \end{center}
    \caption{Average energies~\eqref{eq:average_energy} for $N=101,\,201,\,301$ computed in \cite{chorin91} using a variant of the pivot algorithm. Figure reproduced from \cite{chorin91}.}
    \label{fig:chorin_energy}
\end{figure}

\begin{figure}
    \includegraphics[width=0.75\textwidth]{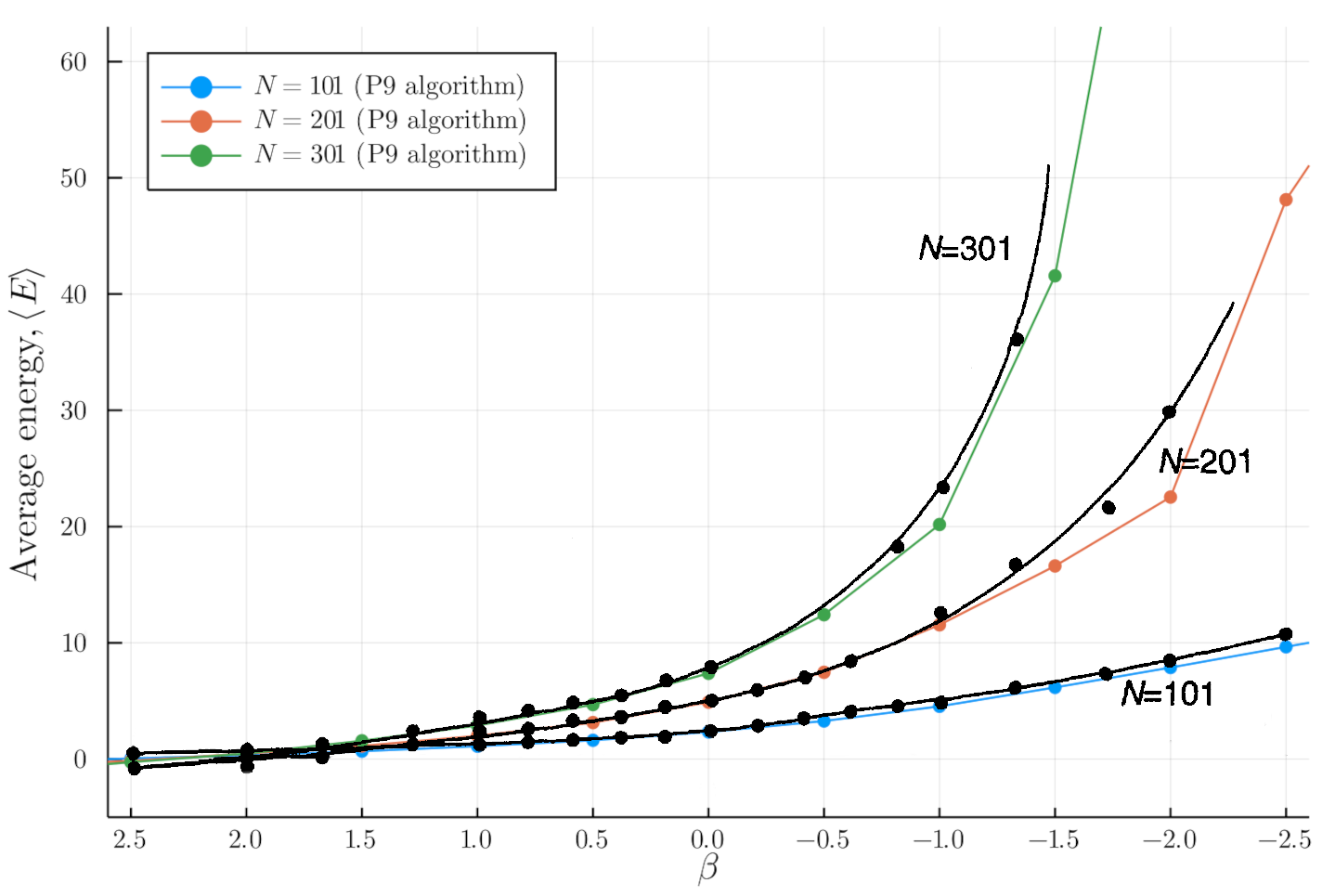}
    \caption{Overlay of Fig.~\ref{fig:chorin_energy} and average energies~\eqref{eq:average_energy} for $N=101,\,201,\,301$ computed using the reduced pivot (P9) algorithm. We observe good agreement between the results in~\cite{chorin91} (black dots) and the pivot algorithm results (color) for $-0.5\le\beta\le1.5$ and hints of disagreement outside of this range.}
    \label{fig:results_overlay_101-301}
\end{figure}

First, consider the results shown in Fig.~\ref{fig:chorin_energy}, which shows the computational results obtained in~\cite{chorin91} for filaments of length $N=101$, $201$, and $301$ steps using a variant of the pivot algorithm. The starting configuration in these computations was chosen to be the straight filament, then $N$ Monte Carlo iterations were performed ``{\it just to begin the process of forgetting the initial conditions}''~\cite{chorin91}, and finally a number of iterations was performed to compute the average energy. We will follow the process described in the previous section, starting from the straight filament at $\beta=-20$ and continuing to increase $\beta$. The results presented in Fig.~\ref{fig:results_overlay_101-301} were generated using the pivot algorithm with the reduced set of $9$ transformations that still ensures ergodicity, the P9 algorithm. The computed results are overlaid by the results from Fig.~\ref{fig:chorin_energy} for ease of comparison. Notice the qualitative similarity of the results and the quantitative agreement for $-0.5\le\beta\le1.5$, although there are some quantitative differences near the endpoints of the displayed interval. Since we do not have the exact details of how the results in~\cite{chorin91} were computed, we cannot address this difference.

\begin{figure}
    \includegraphics[width=0.49\textwidth]{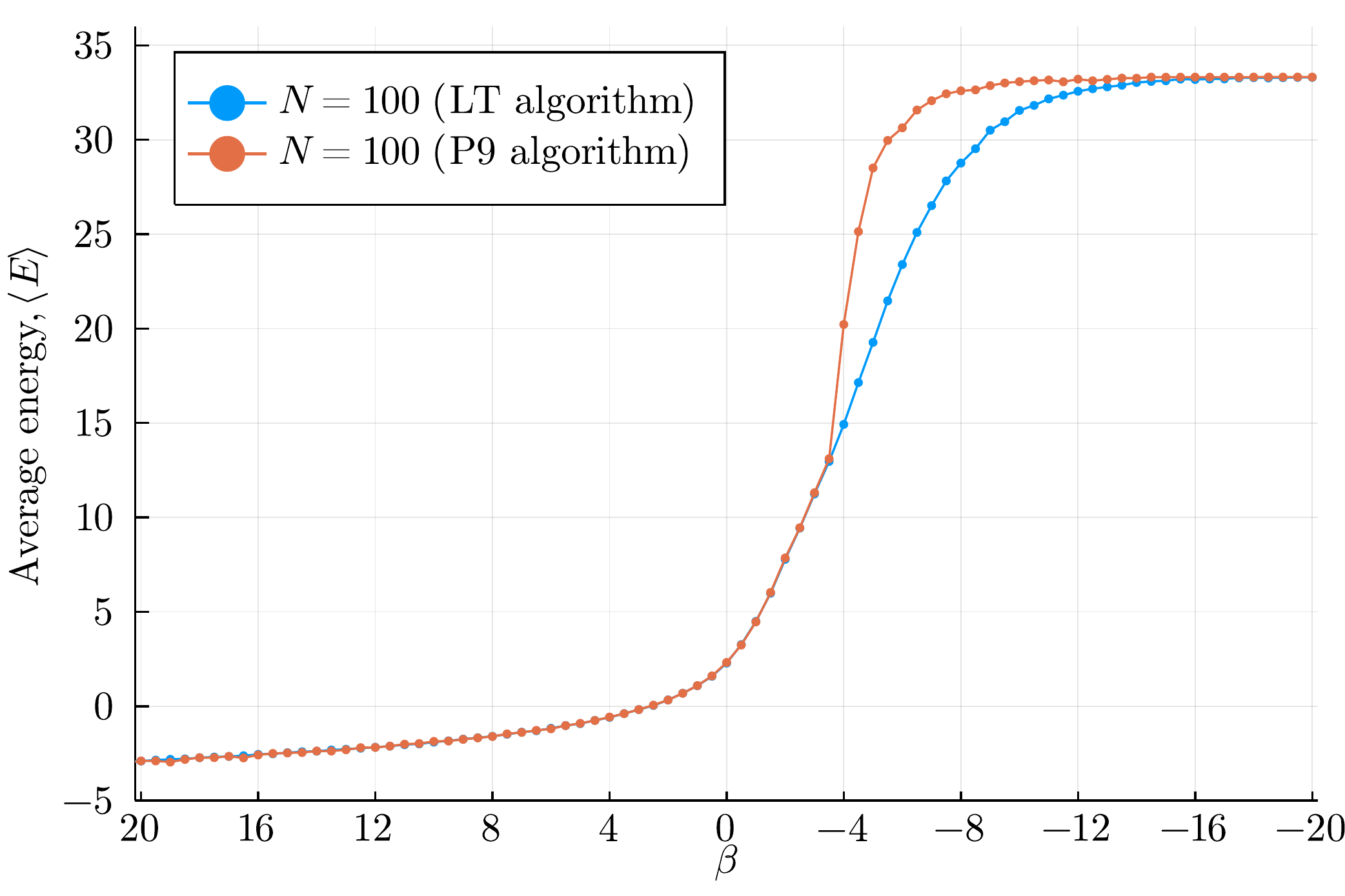}\hfill
    \includegraphics[width=0.49\textwidth]{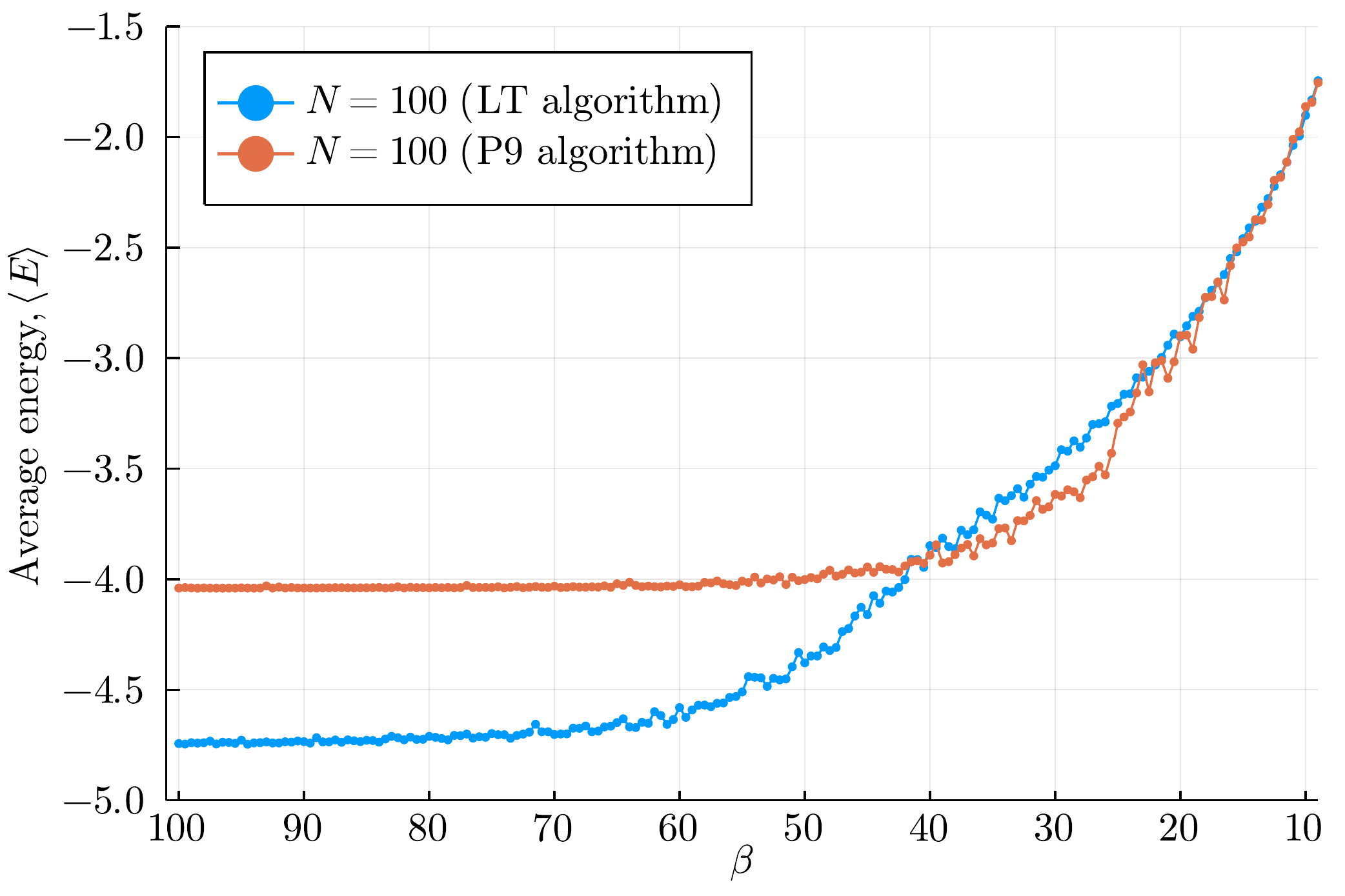}
    \includegraphics[width=0.49\textwidth]{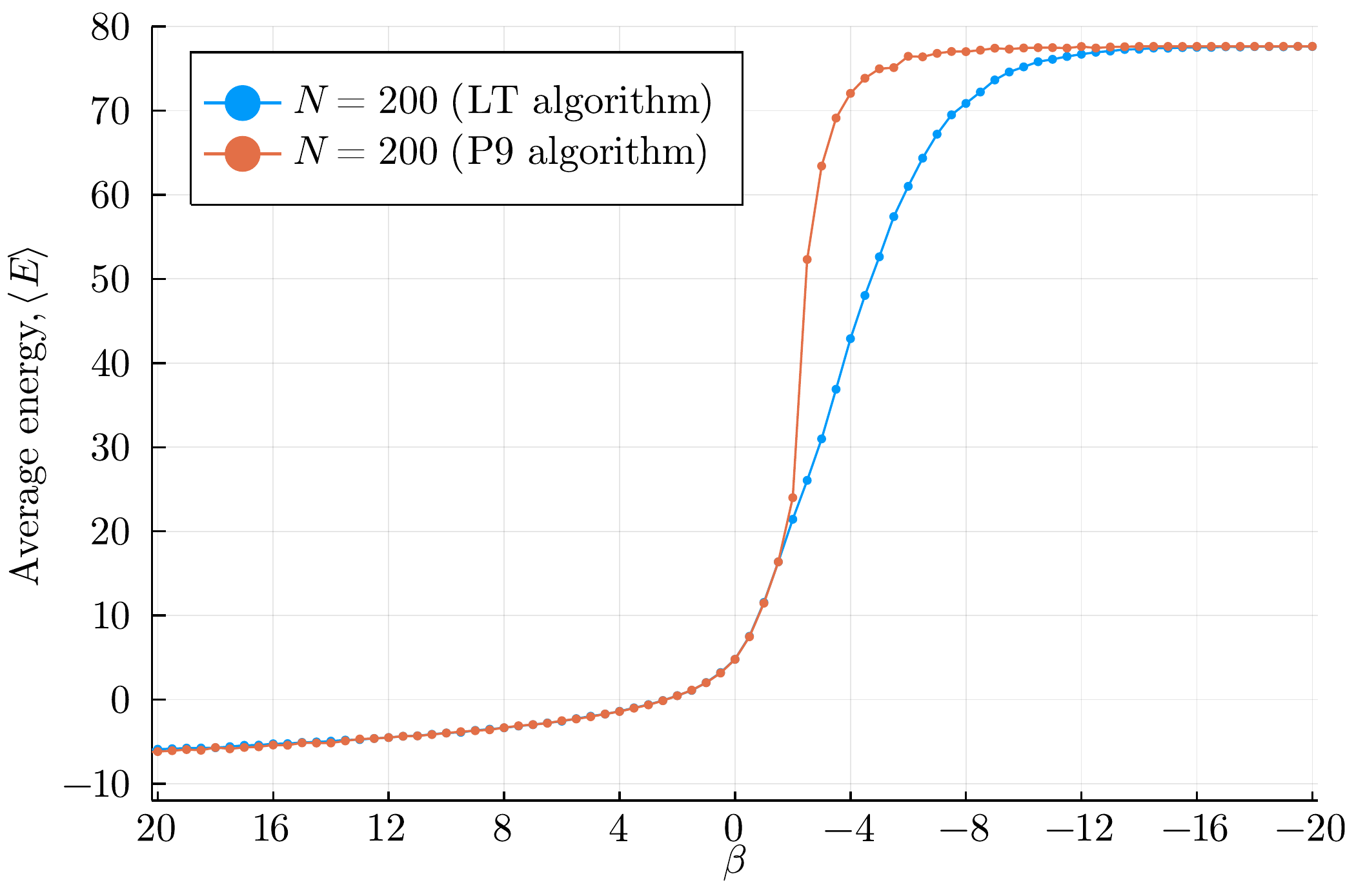}\hfill
    \includegraphics[width=0.49\textwidth]{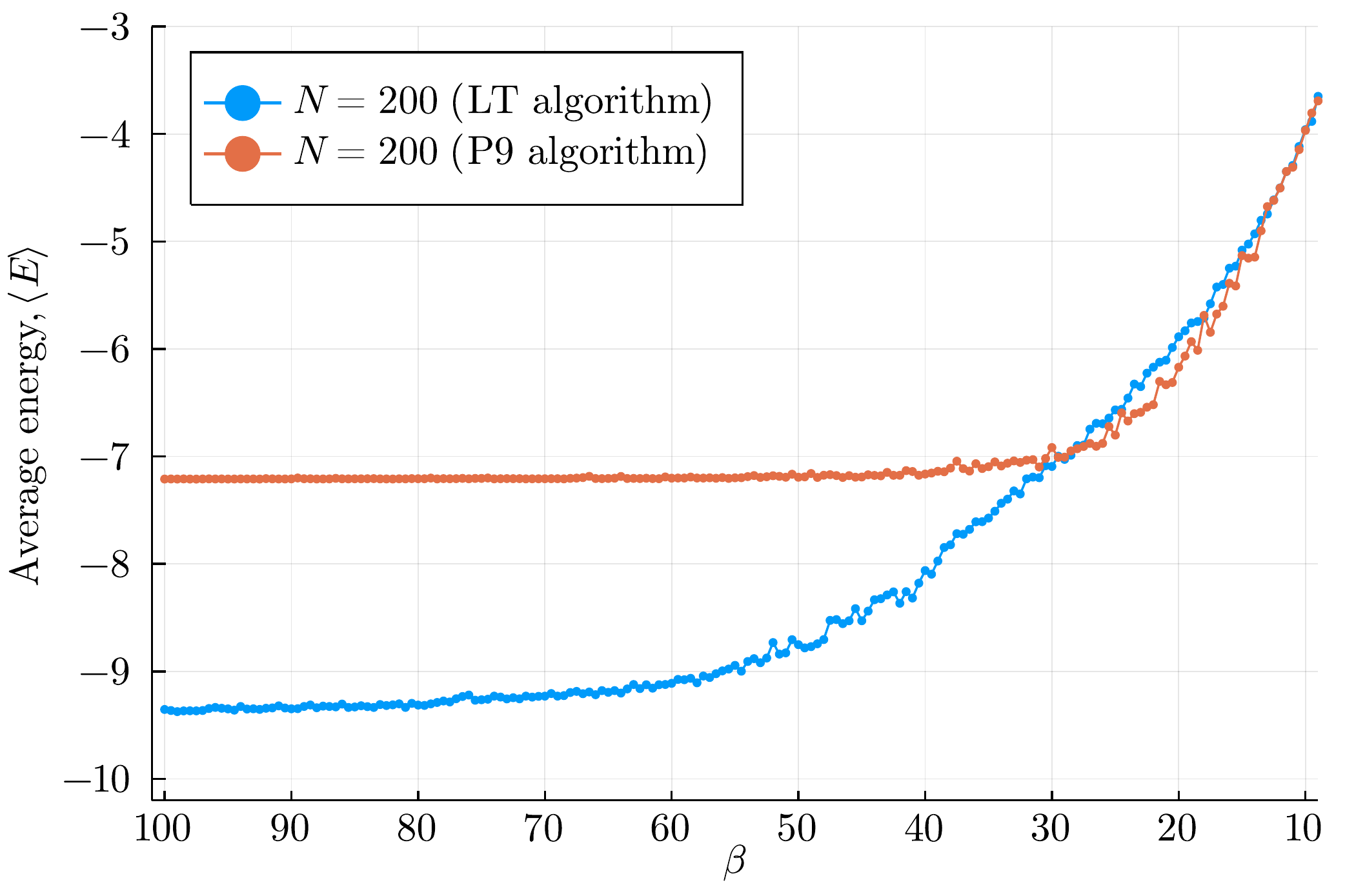}
    \includegraphics[width=0.49\textwidth]{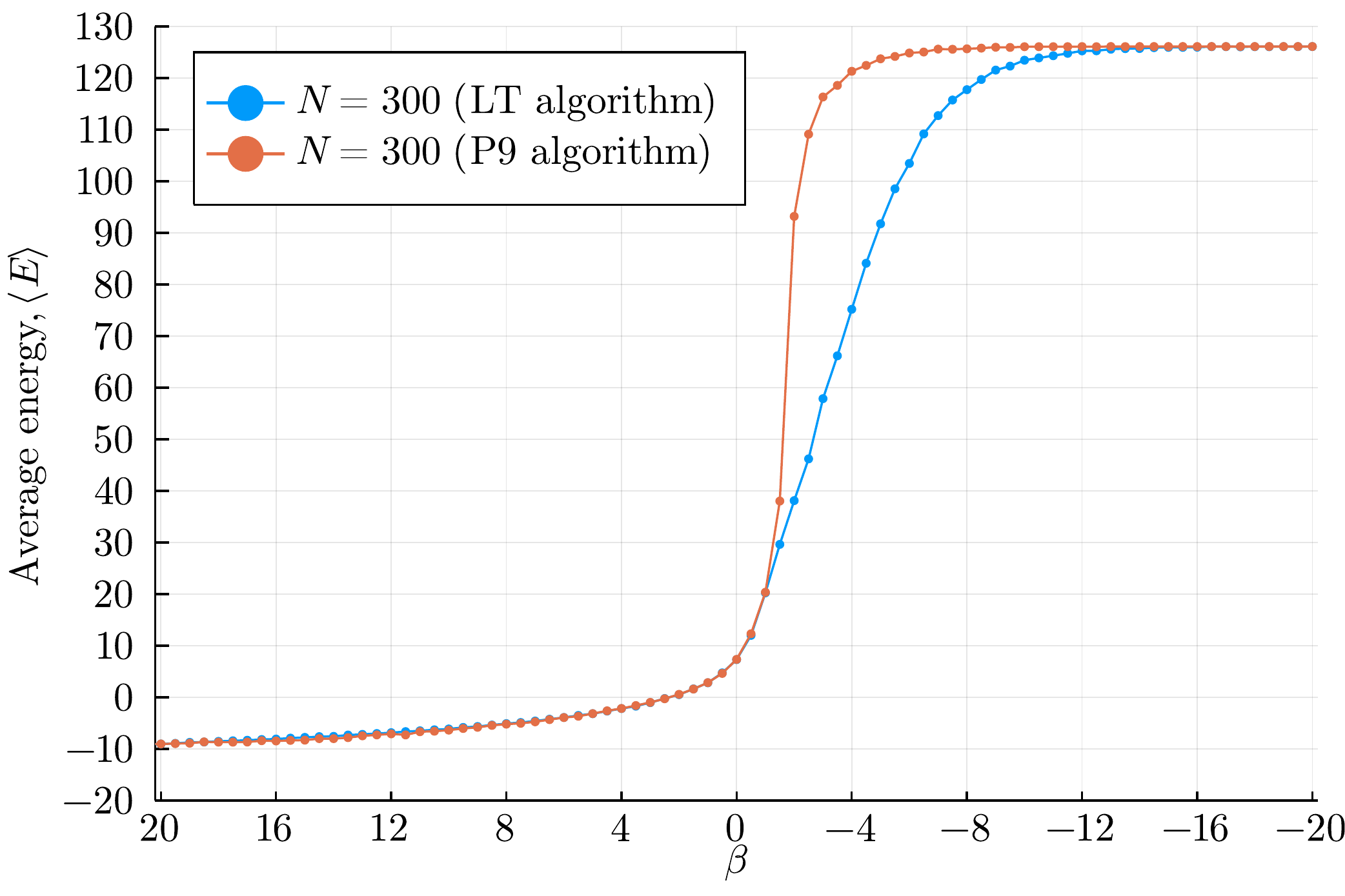}\hfill
    \includegraphics[width=0.49\textwidth]{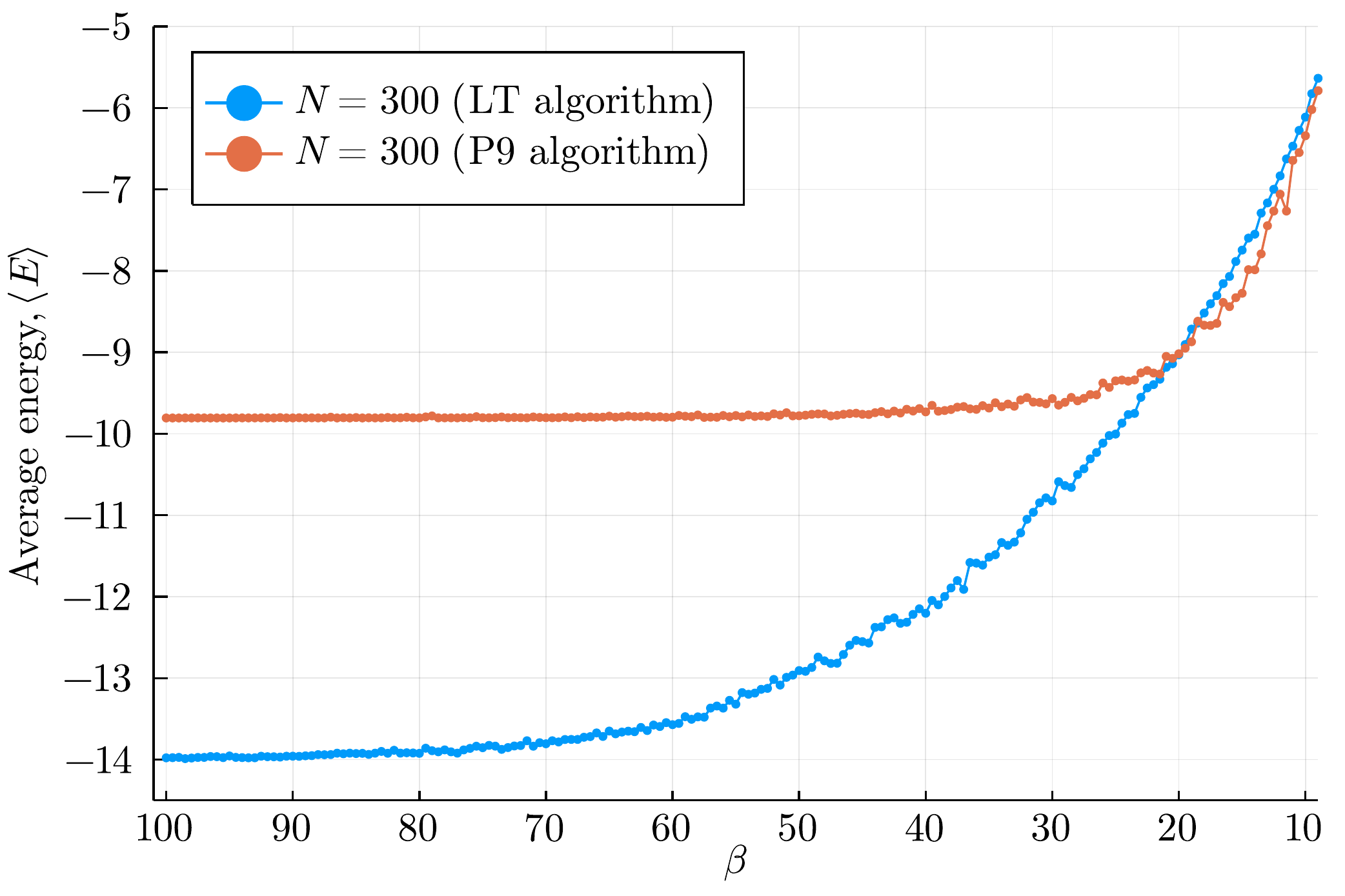}
    \caption{Comparison of the computed average energies with the P9 algorithm (orange) and the LT algorithm (blue). Results are generated using continuation in $\beta$ as described in section~\ref{sec:localized_results}. The pivot algorithm results are generally larger and not reasonable.}
    \label{fig:comparison_LT_P9}
\end{figure}

Next, we use the P9 algorithm and perform the computational regime described in section~\ref{sec:localized_results}. We will start with a straight configuration at $\beta=-20$ and compute average energies while increasing the value of $\beta$ with increments of $\Delta\beta=0.5$. The results for $N=100$, $200$, and $300$ are shown in Fig.~\ref{fig:comparison_LT_P9}. In the left column we show the results of the two algorithms for $-20\le\beta\le20$, and in the right column we show the same for $10\le\beta\le100$. The results from the P9 algorithm are in orange and are generally higher than those from the LT algorithm, which are shown in blue. Notice that in these results the agreement is generally good for $0\le\beta\le10$. Outside this interval we see major differences. For negative values of $\beta$ the results from the pivot algorithm demonstrate the difficulties with moving away from the straight configuration. The average energy stays high for $-10\le\beta\le0$ until, for some small enough value of $\beta$, it ``re-joins'' the results of the LT algorithm. Similarly, there are differences for large positive values of $\beta$. In this case the pivot algorithm gets ``stuck'' in configurations that are far from the minimum energy configurations, and the differences grow as $N$ increases. As pointed out earlier, a discussion of possible minimum energies is in section~\ref{sec:min_energy}.

In order to further demonstrate the difficulties the pivot algorithm will experience for large $|\beta|$, we perform the following computations. Starting with a straight filament, we cycle through the values $\beta=0$, $\beta=100$, $\beta=0$, and $\beta=-20$ (in this order), performing 500,000 burn-in iterations and 200,000 averaging iterations. We then compute the average energies \eqref{eq:average_energy}, average them over six different runs, and compare them to the results of the continuation in $\beta$ with the LT algorithm described in section~\ref{sec:localized_results}. These results are presented in Table~\ref{tab:cycling}. The last two columns of the table show the average energy values computed using the continuation approach with the LT algorithm. Several observations stand out immediately:
\begin{enumerate}
    \item The average energy values for $\beta=0$ agree well for all approaches. The standard deviations for the computed energies prior to averaging were, approximately, $1.3$, $2.0$, and $2.7$ for $N=100$, $200$, and $300$, respectively, so the small differences are acceptable.
    \item The average energy values for $\beta=100$ show significant differences. If we use the LT(c) results as a benchmark, we see that the LT algorithm results are relatively close, but the results from the P9 algorithm fail to get anywhere near the benchmark values. This suggests that the algorithm is getting stuck in some configurations and cannot find a way to improve given the set of available transformations. This is consistent with our discussion in section~\ref{sec:pivot}.
    \item The values for $\beta=-20$ again show significant differences. With this value of $\beta$, the preferred configurations should be very close to straight and so the average energies should be close to the maximum values given in~\eqref{eq:max_energy}. We see that this is achieved for the LT algorithm but by far it is not achieved by the P9 algorithm. For each run, we also recorded the largest computed energy. The LT algorithm reached the straight configuration (maximum energy) in all $18$ runs presented in the table ($3$ values of $N$ and $6$ runs with each $N$). The P9 algorithm achieved the following maximum energy values (up to $4$ significant digits): $E=23.14$ for $N=100$, $E=41.39$ for $N=200$, and $E=60.82$ for $N=300$. The expected values should be near the values listed in the second column of Table~\ref{tab:cycling} labeled ``Straight'', i.e., near $33.32$, $77.64$, and $126.1$, respectively.
\end{enumerate}
\begin{table}
    \begin{tabular}{c|ccccc|cc}
        Algorithm & Straight & $\beta=0$ & $\beta=100$ & $\beta=0$ & $\beta=-20$ & LT(c) $\beta=0$ & LT(c) $\beta=100$ \\
        \hline
        LT, $N=100$ & $33.32$ & $2.275$ & $-4.276$ & $2.289$ & $33.30$ & $2.285$ & $-4.743$ \\
        P9, $N=100$ & $33.32$ & $2.290$ & $-2.779$ & $2.295$ & $22.03$ & $2.285$ & $-4.743$ \\
        LT, $N=200$ & $77.64$ & $4.933$ & $-8.589$ & $4.862$ & $77.62$ & $4.794$ & $-9.355$ \\
        P9, $N=200$ & $77.64$ & $4.789$ & $-3.513$ & $4.808$ & $36.54$ & $4.794$ & $-9.355$ \\
        LT, $N=300$ & $126.1$ & $7.620$ & $-12.50$ & $7.376$ & $126.1$ & $7.328$ & $-13.98$ \\
        P9, $N=300$ & $126.1$ & $7.446$ & $-3.149$ & $7.376$ & $52.14$ & $7.328$ & $-13.98$ \\
    \end{tabular}
    \vskip0.5\baselineskip
    \caption{Results of computing average energies when starting from a straight filament and cycling through the values $\beta=0$, $\beta=100$, $\beta=0$, and $\beta=-20$. Here, LT stands for the algorithm with localized transformations, P9 for the reduced pivot algorithm with $9$ transformations, and LT(c) for the continuation algorithm with localized transformations.}
    \label{tab:cycling}
\end{table}

\subsection{Validation of the LT algorithm}
Without knowing the exact values of the average energy~\eqref{eq:average_energy}, it is impossible to assess the accuracy of the computed results. Due to the growth rate of the number of SAWs as a function of the number of steps $N$ (see Table~\ref{tab:SAWs}), it is only realistic to compare computed results to exact values for small values of $N$. To this end we completely enumerated all SAWs for $N=3,\dots,9$ and obtained exact values of $\langle E\rangle$ for various values of $-100\le\beta\le100$.

\begin{figure}
    \includegraphics[width=0.49\textwidth]{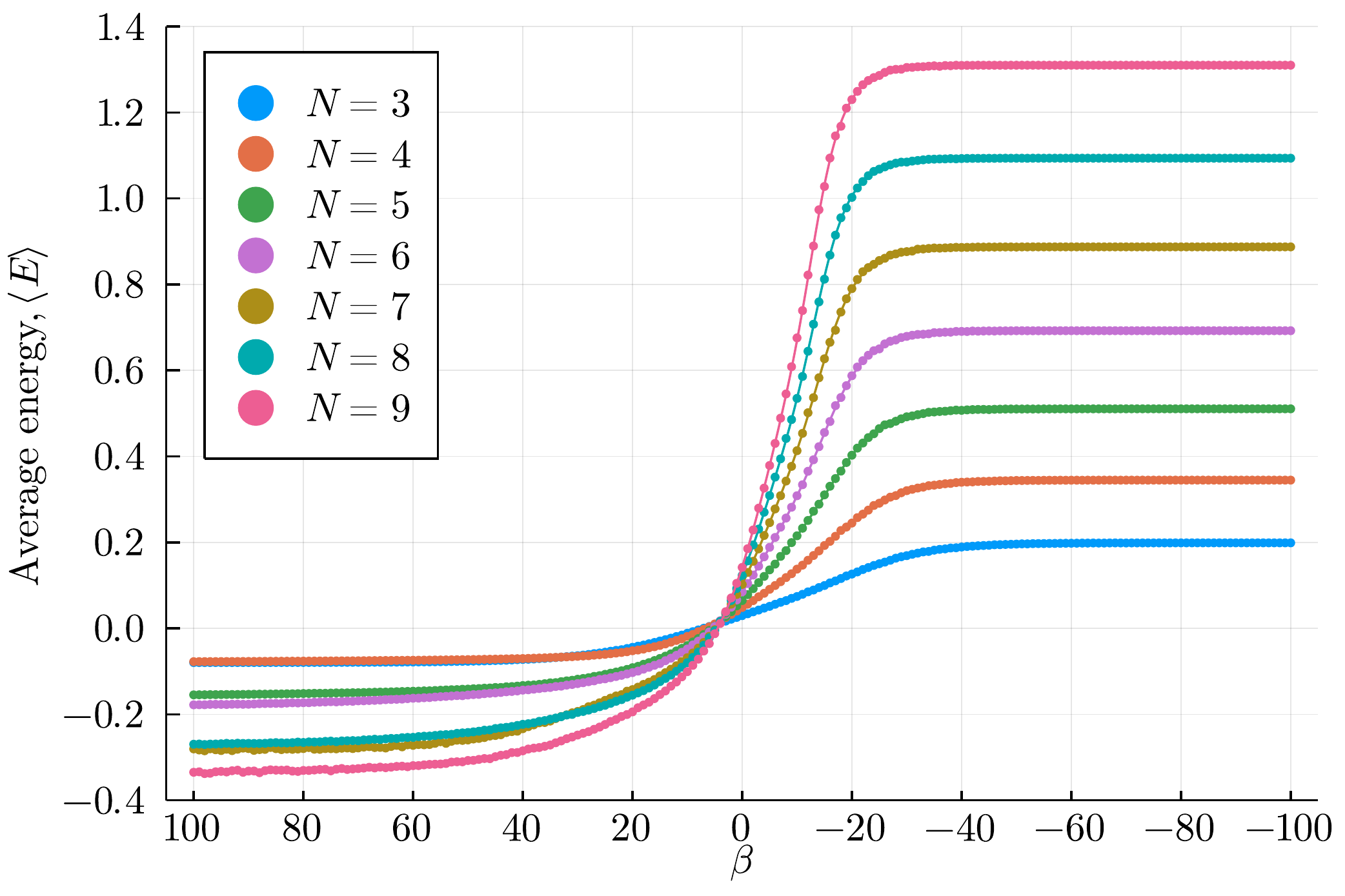}
    \hfill
    \includegraphics[width=0.49\textwidth]{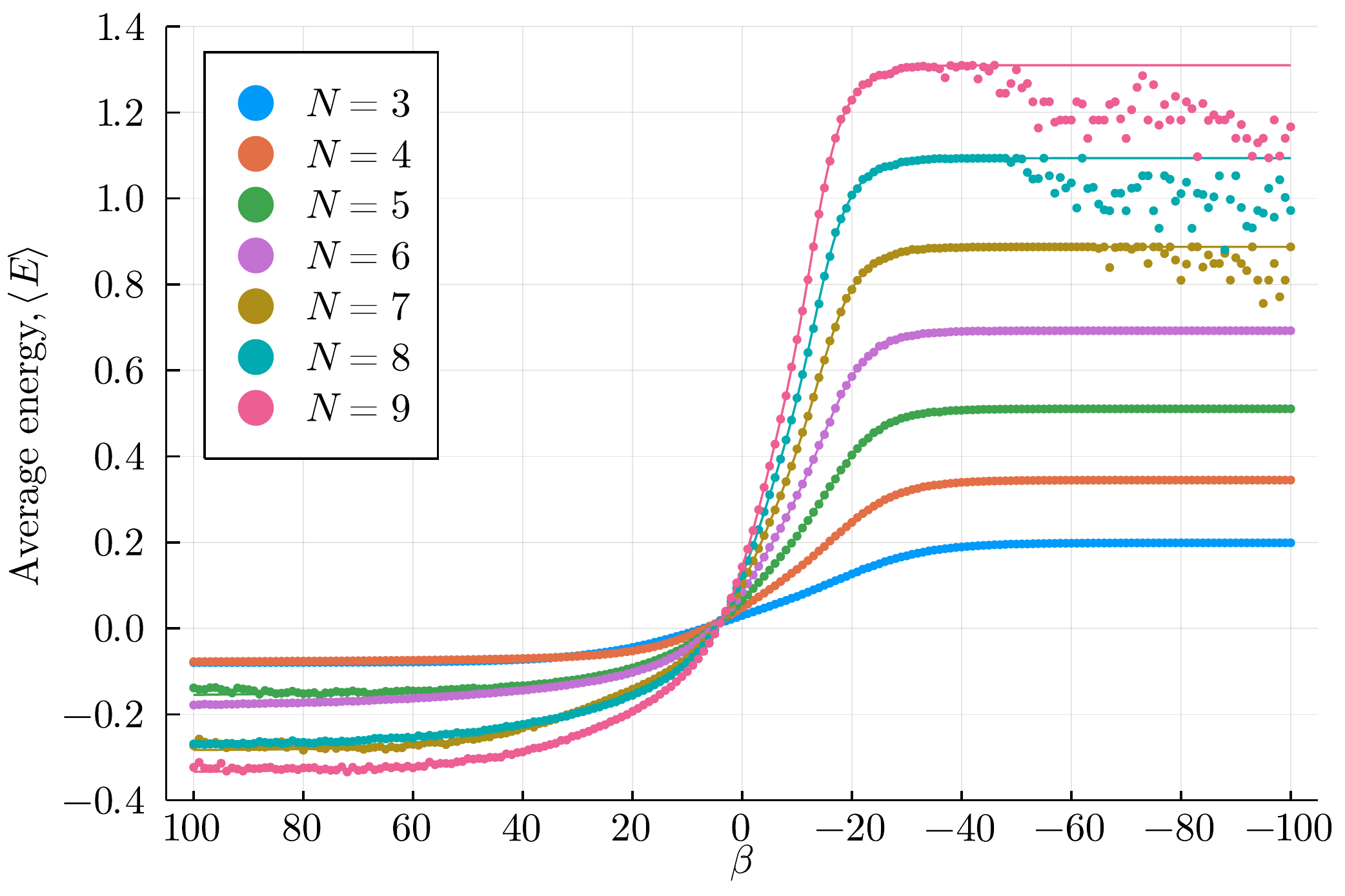}
    \caption{Results for the average energy $\langle E\rangle$ for filaments of lengths $N=3,\dots,9$. Computed results (dots) overlay exact values (solid curve). The plot on the left corresponds to the algorithm with localized transformations, the plot on the right corresponds to the reduced pivot algorithm with $9$ transformations.}
    \label{fig:validation_results}
\end{figure}

To demonstrate the capabilities of the LT algorithm and highlight the difficulties the P9 algorithm will experience, we perform the following test. Given $N$ and $\beta$, start with a straight configuration but first perform $10N$ iterations with $\beta=0$ to get away from it. Then first perform 100,000 iterations with the given $\beta$ to forget history and then perform 200,000 averaging iterations to approximate $\langle E\rangle$ for that $\beta$. Repeat with various values of $\beta$ in the interval of interest. The computed results, together with the exact values, are shown in Fig.~\ref{fig:validation_results}. In the left plot we show the results using the LT algorithm and in the right plot we show the results using the P9 algorithm. The computed approximations of $\langle E\rangle$ for each integer $\beta$ are shown as dots, and the exact values, computed with increments $\Delta\beta=0.1$, are shown as curves. Notice how the results with the LT algorithm (left plot) agree well with the exact values, showing that the algorithm is flexible enough to handle both positive and negative values of $\beta$. On the other hand, the P9 algorithm struggles to recover the straight configurations for $N=7$, $8$, and $9$ (seen in the top right part of the right plot) even with 300,000 iterations, and there are also signs of struggle to approximate well the lowest energy configurations (seen in the bottom left part of the right plot). Similar struggles, but to a much lesser degree, are also showing in the left plot. Overall, both algorithms perform well for $-30\le\beta\le30$ for these values of $N$.

\subsection{Entropy approximations}
\begin{figure}
  \begin{center}
    \includegraphics[width=0.68\textwidth]{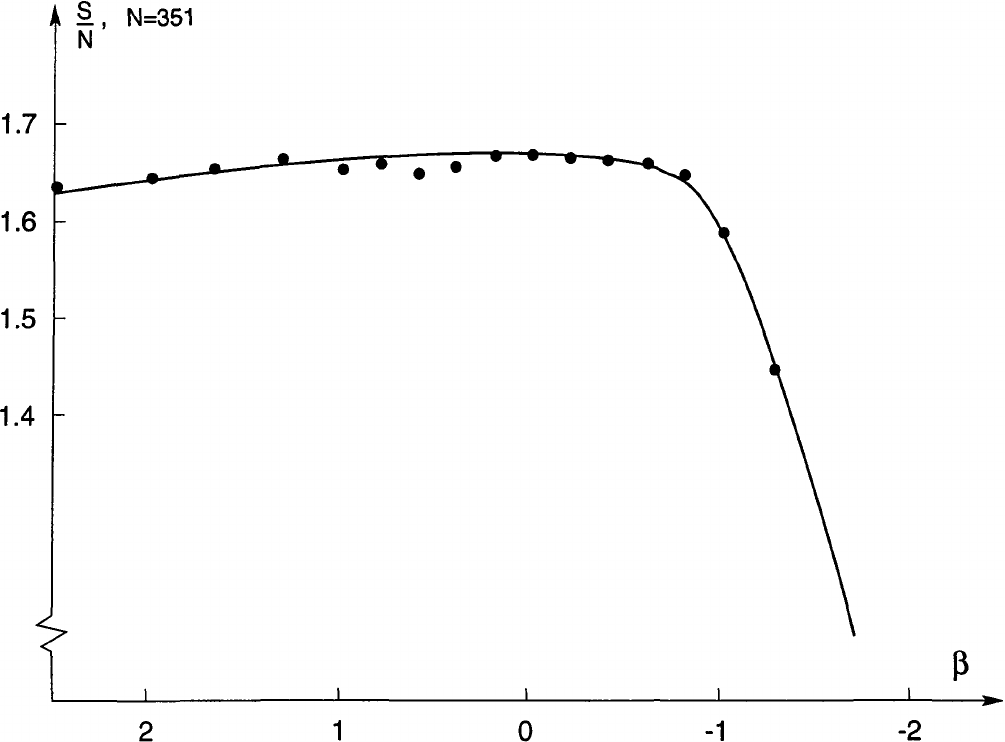}
  \end{center}
  \caption{The computational results of entropy of a vortex filament per unit length, $S/N$, as a function of $\beta$ for a filament of length $N=351$ \cite{chorin91,chorin}. Figure reproduced from \cite{chorin91}.}
  \label{fig:chorin_entropy}
\end{figure}

In Chorin's work \cite{chorin91,chorin}, entropy of the system at various temperatures is estimated using an algorithm based on an earlier work by Meirovitch \cite{meirovitch83}. The idea is to enumerate all possible very short SAWs ($150$ SAWs of length $3$ are used in \cite{chorin91}) and then use their relative frequencies in a Monte Carlo run to approximate the probabilities needed to evaluate an entropy expression similar to \eqref{eq:entropy}. A result for entropy per unit length of the ensemble of vortex filaments, $S/N$, as a function of $-1.5\le\beta\le2.5$, is shown in Figure~\ref{fig:chorin_entropy}, where $N=351$ is used. It is observed in \cite{chorin91} that {\it ``$S$ has a maximum at $T=\infty$ ($\beta=0$), as expected. The slope of $S$ is much smaller on the positive $T$ side than on the negative $T$ side, as can be expected from the larger values of $E$ for $T<0$ and from the relation $T^{-1}=\partial S/\partial E$. Further, note that $S/N$ varies little with $N$ in the range where the calculation can be trusted, and thus $S$ increases with $N$. The larger the filament, the larger its entropy.''}

\begin{figure}
  \begin{center}
    \includegraphics[width=0.75\textwidth]{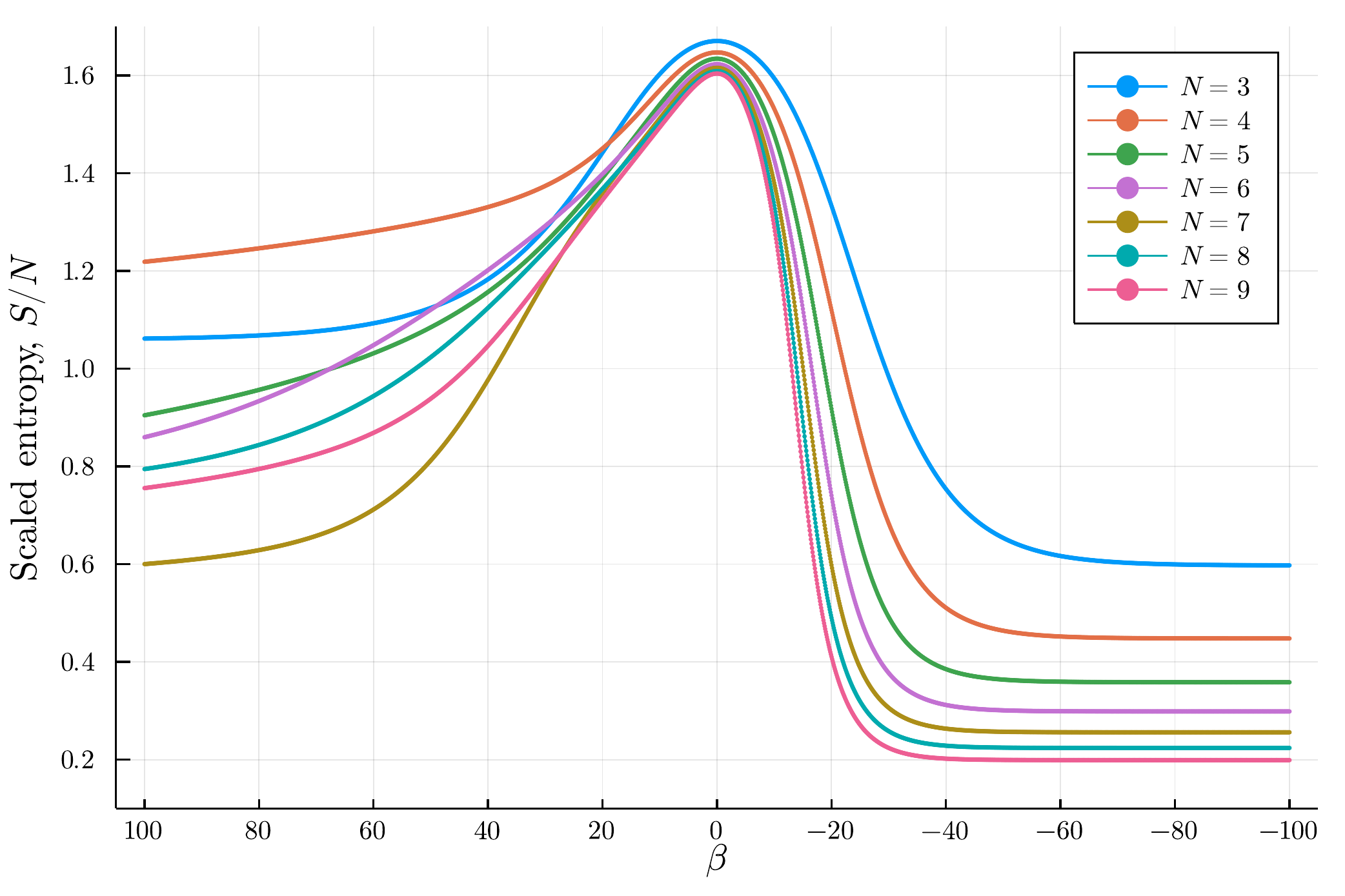}
  \end{center}
  \caption{Exact values of entropy per unit length for $N=3,\dots,9$. The curve ordering is best seen for large negative $\beta$ values where $S/N$ decreases as $N$ increases.}
  \label{fig:entropy_3-9}
\end{figure}

In order to validate these results and extend them to a larger range of $\beta$, we used as benchmarks the cases for which we had the complete enumerations ($N=3,\dots,9$). The exact entropies for these cases are shown in Figure~\ref{fig:entropy_3-9}. As discussed in section~\ref{sec:statmech}, for cases where exact enumeration is possible the entropy for $\beta=0$ is easy to evaluate as $S=\log{M}$, where $M$ is the number of SAWs of a given length. Self-avoiding walks on a cubic lattice have been enumerated for lengths up to $N=36$ \cite{sbb11}, and this data can be used to construct approximate expressions $\tilde{M}$ for $M$ as a function of $N$ for $N>36$. While the reference \cite{sbb11} is missing one parameter value ($c_1$), its {\tt arXiv.org} version~\cite{sbb_arxiv} provides a slightly different, but completely described formula to approximate $M$. For completeness, we reproduce it here,
\begin{equation}
  M
  \approx
  \tilde{M}
  =
  A\mu^NN^\theta(1+cN^{-\Delta}+k(-1)^NN^{-\alpha}),
  \label{eq:formula_for_M}
\end{equation}
where $A=1.1951966888$, $\mu=4.6840041570$, $\theta=0.1597395125$, $c=0.1227360755$, $\Delta=1.4315024046$, $k=-0.0619076482$, and $\alpha=1.8985141134$. We note that the relative errors resulting from this formula for $N=14,\dots,36$ are all below $2\times10^{-6}$; this results in absolute errors for the entropy, $\log\tilde{M}-\log M$, to be of the same order. Consequently, the resulting approximation
\begin{equation}
  S/N
  =
  (\log M)/N
  \approx
  (\log\tilde M)/N
  \label{eq:entropy_approx}
\end{equation}
is expected to be quite good and perhaps improve for longer filaments due to the $N$ in the denominator.

Using \eqref{eq:formula_for_M} and \eqref{eq:entropy_approx} to approximate $S/N$ for $N=351$, we obtain $S/N\approx1.54733$, which appears significantly below the maximum value of about $1.67$ estimated from Figure~\ref{fig:chorin_entropy}. In fact, as $N\to\infty$, the quantity $(\log\tilde M)/N$ eventually monotonically decreases and approaches $\log\mu\approx1.54415$. Note that even for $N=9$ we have $S/N=(\log{1853886})/9\approx1.60364$. Clearly, a better algorithm for computing entropy than the one used in \cite{chorin91} is needed.

Improvements of the Meirovitch algorithm have appeared in the literature more recently and have been applied to magnetic systems, polymers, peptides, and liquids (see, e.g., \cite{cheluvarajameirovitch05,whitefuntmeirovitch05,whitemeirovitch06} and references within). A possible promising approach would be to use a variant named the {\it hypothetical scanning Monte Carlo} (HSMC) method, which approximates the probability, $p(x)$, of a given filament configuration, $x$, of length $N$ in the configuration space $\mathcal{S}_N$ at a given $\beta$. We provide here a brief sketch of the algorithm, which is described more fully in the references.

The approximation to $p(x)$, denoted by $p^\text{HSMC}(x)$, is computed as a product of conditional ``transition'' probabilities that the $k$th step of the filament is in the given position as in $x$, given that the previous $k-1$ steps are fixed as in $x$. To generate these probabilities, we would keep the first $k-1$ steps fixed and apply the LT algorithm as used to compute $\langle E\rangle$ by allowing reconstructions only between the nodes $k$ through $N$. The transition probability would then be approximated by the ratio of the filaments that agree with the $k$th step of $x$ versus all filaments generated in the Monte Carlo run. Clearly, this process would be computationally intensive, since $N-1$ Monte Carlo runs (conceivably shorter and shorter) would have to be performed to compute a single $p^\text{HSMC}(x)$. On the other hand, $p(x)$ could be approximated arbitrarily closely by allowing sufficiently long simulations.

With the computed $p^\text{HSMC}(x)$ the Helmholtz free energy, $F$, could be approximated via \eqref{eq:helmholtz},
\begin{equation*}
  F
  \approx
  F^\text{HSMC}
  =
  E(x)+\frac{1}{\beta}\log{p^\text{HSMC}(x)},
\end{equation*}
where $E(x)$ is the easily computed energy of the filament $x$. Subsequently, the entropy could be approximated through the relationship $F=\langle E\rangle-TS$ if the average energy has been computed, as shown in \eqref{eq:entropy3},
\begin{equation*}
  S
  \approx
  S^\text{HSMC}
  =
  k_B\beta(\langle E\rangle-F^\text{HSMC})
  =
  k_B(\beta\langle E\rangle-\beta E(x)-\log{p^\text{HSMC}(x)}).
\end{equation*}
It follows that the error in the entropy computation, $|S-S^\text{HSMC}|$, will be affected both by the accuracy of the HSMC computation of $p^\text{HSMC}(x)$ as well as the accuracy of the computed $\langle E\rangle$; more precisely, it will be roughly bounded by the sum of the relative error in approximating $p(x)$ and $|\beta|$ times the absolute error in approximating $\langle E\rangle$ (all multiplied by $k_B$).

Note that to compute the entropy of the system of filaments of length $N$ for many possible values of $\beta$ (such as shown in Fig.~\ref{fig:entropy_3-9} for example) requires $N-1$ Monte Carlo simulations for each value of $\beta$ and thus a significant amount of CPU time. We have performed such computations for $N=3,\dots,9$ for validation purposes, but since in the next section we will present an alternative, and much more efficient, approach, we will not present the results here.

\subsection{Entropy computations based on energy values}
A very efficient alternative to computing the entropy can be used if good approximations to $\langle E\rangle$ have been computed across an interval of $\beta$ values and a single value for entropy is known. Recall the relationship \eqref{eq:dSdE}, restated here for convenience:
\begin{equation}
  \frac{\partial S}{\partial\langle E\rangle}
  =
  k_B\beta.
  \label{eq:dSdE2}
\end{equation}
If the $\beta$ interval is partitioned with values $\beta_i$, the corresponding (average) energies are denoted by $E_i$, and the (sought) entropies by $S_i$, then, provided $S$ is smooth enough, \eqref{eq:dSdE2} can be discretized as
\begin{equation*}
  \frac{S_{i+1}-S_i}{E_{i+1}-E_i}
  =
  k_B\beta_{i+1/2}
  +
  \frac{(\Delta\beta_i)^3}{48}\left.\frac{\partial^3S}{\partial\langle E\rangle^3}\right|_{\langle E\rangle=\tilde e}
\end{equation*}
for some $\tilde e$ between $E_i$ and $E_{i+1}$, where $\beta_{i+1/2}=(\beta_{i+1}+\beta_i)/2$ and $\Delta\beta_i=|\beta_{i+1}-\beta_i|$. This then leads to the following Euler's method for approximating the entropy,
\begin{equation}
  S_{i+1}
  =
  S_i
  +
  k_B\beta_{i+1/2}(E_{i+1}-E_i),
  \label{eq:entropy_stepping}
\end{equation}
provided at least one of the values $S_i$ is known to start the recursion. Since, as discussed above, the entropy when $\beta=0$ can be computed for small $N$ or approximated for larger $N$ by $k_B\log{\tilde M}$ using \eqref{eq:formula_for_M}, a starting value is readily available. Alternatively, one can use the HSMC algorithm to approximate the entropy for any one value of $\beta$, for example $\beta=0$.

\begin{figure}
  \begin{center}
    \includegraphics[width=0.75\textwidth]{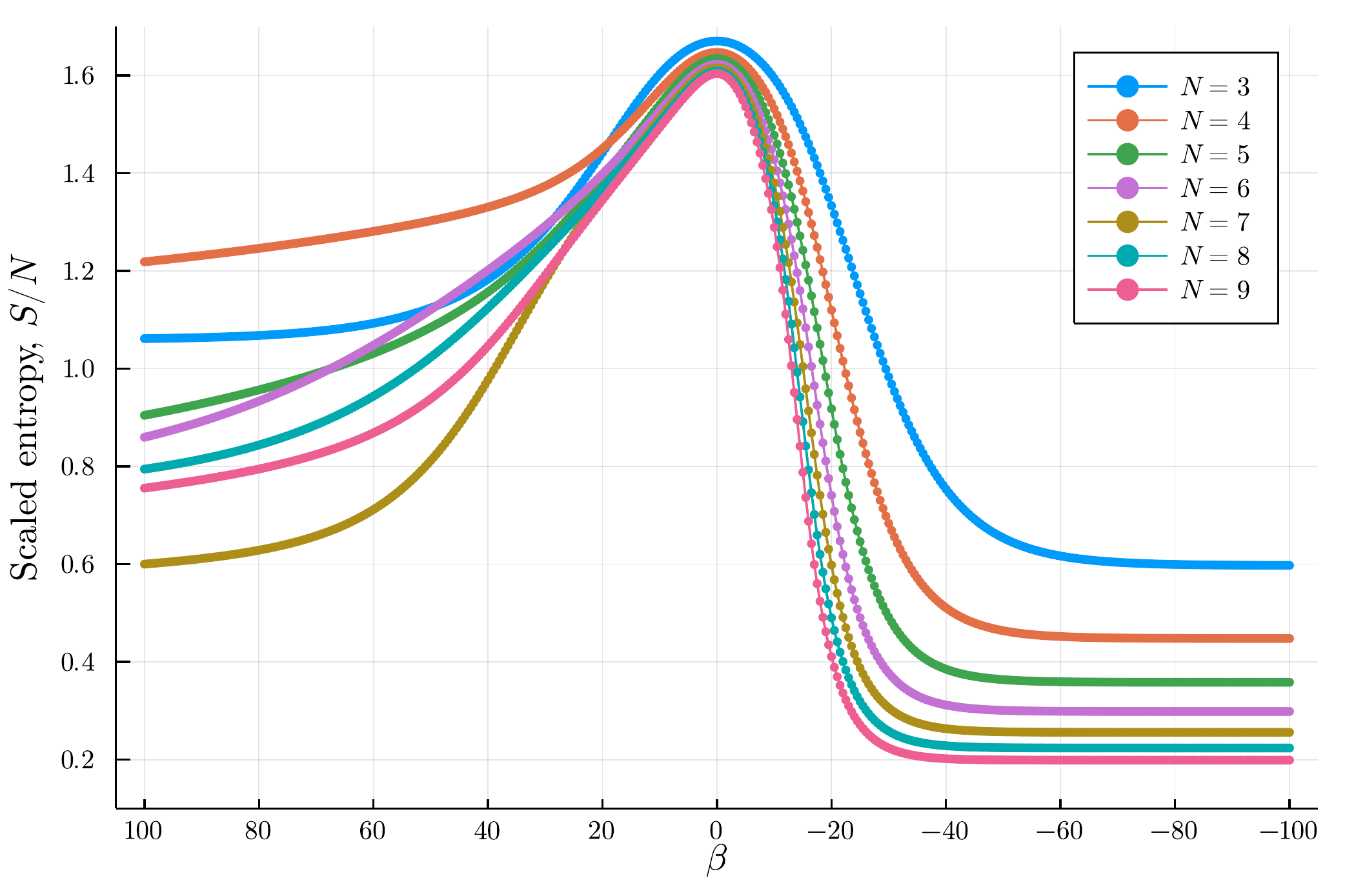}
  \end{center}
  \caption{Approximate values of entropy per unit length for $N=3,\dots,9$ computed using the algorithm~\eqref{eq:entropy_stepping}, exact energy values, and $\Delta\beta=0.5$. Approximate values (large dots) overlay exact values (curves).}
  \label{fig:entropy_stepping_exact_energy_3_9}
\end{figure}

To validate this approach, we first apply it to the cases with $N=3,\dots,9$ and with exact (average) energy values provided at equidistant values of $\beta$ with $\Delta\beta=0.5$. The results are shown in Figure~\ref{fig:entropy_stepping_exact_energy_3_9}. The exact value of entropy at $\beta=0$ is taken as $\log M$ with $M$ from Table~\ref{tab:SAWs}. The Euler algorithm \eqref{eq:entropy_stepping} is applied to compute the remaining entropy values. As shown in the figure, the computed values (large dots) agree well with the exact values (underlying curves). We also see that even after $100$ steps in each direction away from $\beta=0$ the agreement is excellent.

\begin{figure}
  \begin{center}
    \includegraphics[width=0.75\textwidth]{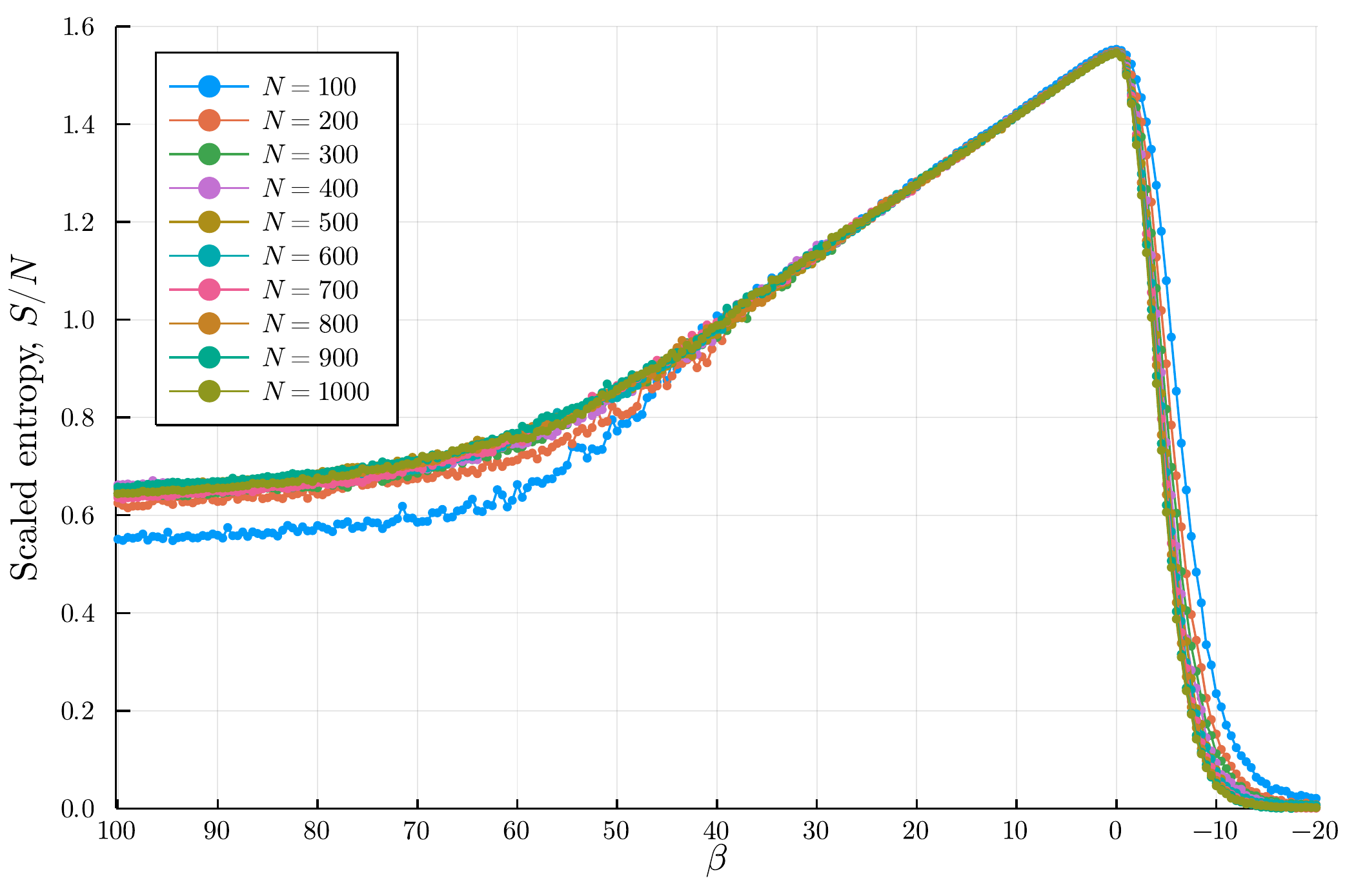}
  \end{center}
  \caption{Approximate values of entropy per unit length for $N=100,\dots,1000$ computed using the algorithm~\eqref{eq:entropy_stepping}, values of $\langle E\rangle$ from the continuation LT algorithm, and $\Delta\beta=0.5$.}
  \label{fig:entropy_stepping_100_1000}
\end{figure}

For $N=100$, $200$, \dots, $1000$, we use \eqref{eq:formula_for_M} and \eqref{eq:entropy_approx} to get an approximation for the entropy at $\beta=0$, and apply algorithm~\eqref{eq:entropy_stepping} with the precomputed average energy values. For the values of $\langle E\rangle$ shown in Fig.~\ref{fig:results-100-1000}, the results for the scaled entropy, $S/N$, are shown in Figure~\ref{fig:entropy_stepping_100_1000}. Here, $\Delta\beta=0.5$ as in the validation case shown in Fig.~\ref{fig:entropy_stepping_exact_energy_3_9}. As expected, the scaled entropy has a maximum value when $\beta=0$ and the rate of increase for $\beta>0$ is smaller (in absolute value) than the rate of decrease for $\beta<0$. Notice how the proximity of the ten curves suggests that entropy $S$ increases linearly with the length of the filament, $N$. We note that as $\beta\to-\infty$, $S/N\to0$ since there are six straight configurations with highest probability, so $S\to\log{6}$. On the other side, as $\beta\to\infty$, the fact that $S/N$ appears to be approaching a finite limit suggests that the number of energy minimizing configurations grows exponentially as $N\to\infty$.

\begin{figure}
  \begin{center}
    \includegraphics[width=0.75\textwidth]{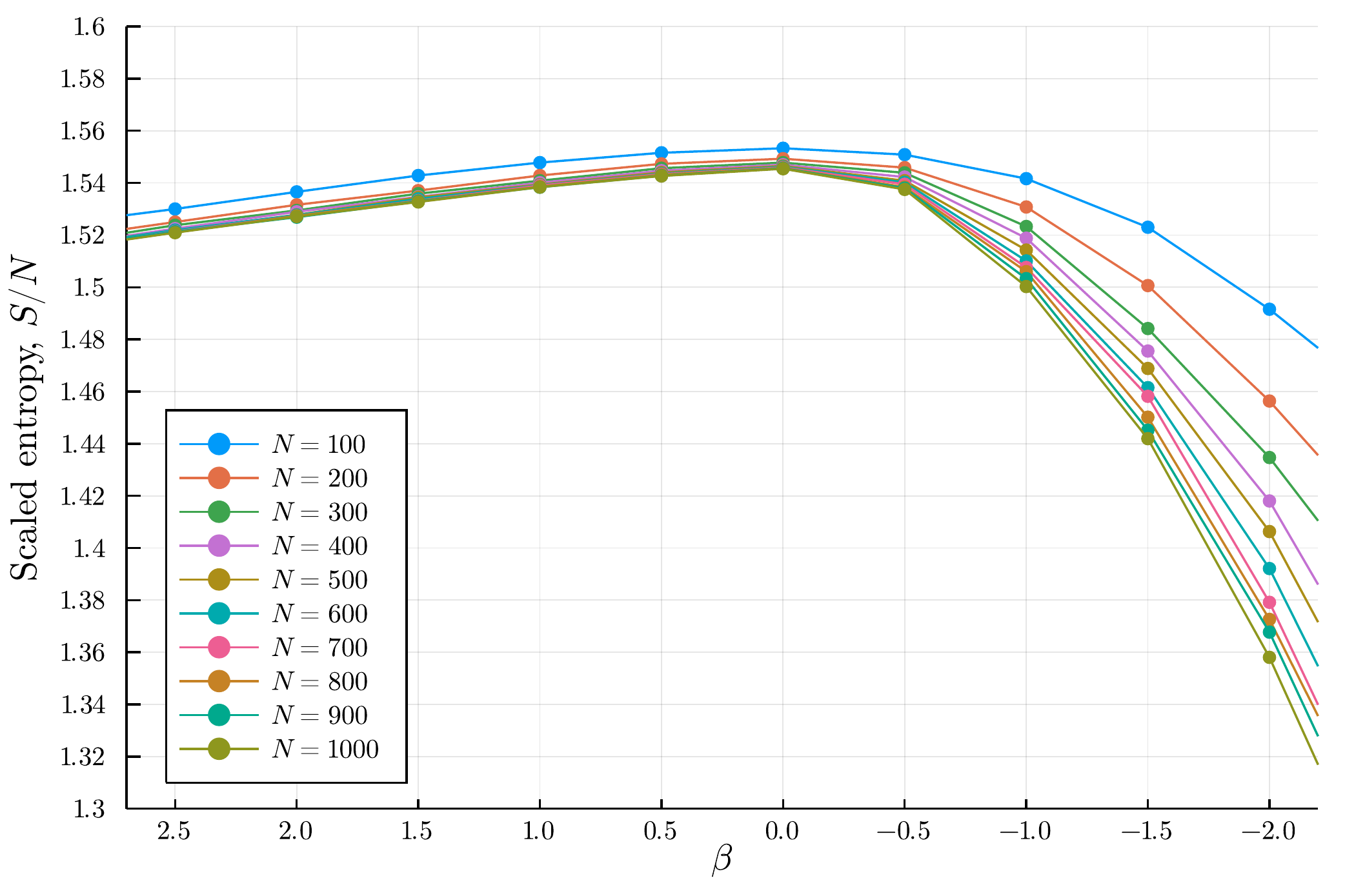}
  \end{center}
  \caption{Same as in Fig.~\ref{fig:entropy_stepping_100_1000} but displayed on a shorter interval for comparison with Fig.~\ref{fig:chorin_entropy}.}
  \label{fig:entropy_stepping_100_1000_short_interval}
\end{figure}

To make a comparison to the results in~\cite{chorin91} and shown in Fig.~\ref{fig:chorin_entropy}, in Fig.~\ref{fig:entropy_stepping_100_1000_short_interval} we show a zoomed-in version of the computed entropies using an interval similar to that in Fig.~\ref{fig:chorin_entropy}. Notice that qualitatively the results are similar. The results in Fig.~\ref{fig:chorin_entropy} are for $N=351$, so our results for this $N$ would appear between the third and the fourth curves from the top. Quantitatively we notice differences in both the slopes on the two sides of the maximum, as well as the maximum value itself.

\subsection{Minimum energy results}
\label{sec:min_energy}
Even though we do not have a direct comparison to exact values of the average energy~\eqref{eq:average_energy}, it appears that the new LT algorithm allows for accurate approximation of the average energies for negative values of $\beta$ and for values close to $0$. How accurate the computed energies for large positive values of $\beta$ are correct is less clear. It follows from \eqref{eq:boltzmann_distribution} that for such values the probabilistically preferred configurations are those with lowest energies, but it is not clear what the lowest possible energy is. This is in contrast with the largest possible energy value that is clearly associated with the straight filament and its value is easily computable via \eqref{eq:max_energy}.

Intuitively, the energy expression~\eqref{eq:energy} suggests that the lowest energy will be associated with a filament folded in such a way that individual steps line up close to each other in an antiparallel way, resulting in large negative contributions to the energy, and also in such a way that there are many right angles which contribute zero energy. From a fluid mechanics point of view this is also intuitive since antiparallel vortices with the same circulation should contribute negligible kinetic energy away from the vortices.

To assess how well the new algorithm with localized transformations can reconstruct the low-energy configurations, we have first computed the lowest possible energies for short filaments. We fully enumerated all filaments of lengths up to $N=18$, and then based on these results we made simplifying assumptions to extend the results up to $N=21$. These results (with $6$ significant digits) are shown in Table~\ref{tab:energies_min}.
\begin{table}
  \begin{tabular}{c|c}
    $ N$ & $E_{\text{min}}(N)$ \\
    \hline
    $ 2$ & $0$  \\
    $ 3$, $4$ & $-1$  \\
    $ 5$ & $-2$  \\
    $ 6$ & $-2.29289$  \\
    $ 7$ & $-3.58579$  \\
    $ 8$ & $-3.45603$  \\
    $ 9$ & $-4.32627$  \\
    $10$ & $-4.86986$  \\
    $11$ & $-5.80649$  \\
    $12$ & $-5.86948$  \\
    $13$ & $-6.94895$  \\
    $14$ & $-7.14629$  \\
    $15$ & $-8.10289$  \\
    $16$ & $-8.37101$  \\
    $17$ & $-9.17890$  \\
    $18$ & $-9.63666$  \\
    $19$ & $-10.6340$  \\
    $20$ & $-10.9049$  \\
    $21$ & $-11.7208$  \\
  \end{tabular}
  \vskip0.5\baselineskip
  \caption{Minimum energy values for filaments of length $N$ on a cubic lattice. Results up to $N=18$ are exact minima; results for $N>18$ are obtained using an assumption based on the empirical observation that the endpoints of the filament are in the same unit square on the cubic lattice.}
  \label{tab:energies_min}
\end{table}
Looking at the limited set of data, it appears to exhibit a strong linear pattern. The best fit line has the equation $E_\text{min}\sim-0.604671N+1.13186$ and $R^2=0.996184$. We then used our numerical approach (LT algorithm) and attempted to locate the minimum-energy configurations for larger values of $N$. Since these have not been validated by another approach, we will not report them here. These computed energies follow the linear trend and when using them to generate a best-fit line, we obtain the equation
\begin{equation}
    E_\text{min}
    \sim
    -0.612418N+1.19011
    \label{eq:best_fit_30}
\end{equation}
with $R^2=0.998585$. Using this slightly steeper line to generate predictions for the minimum energies for $N=100,\dots,1000$, we obtain the results in the column labeled ``Predicted $E_\text{min}$'' in Table~\ref{tab:energies_min_100_1000}.
\begin{table}
  \begin{tabular}{c|c|cc|cc}
    $ N$ & Predicted $E_{\text{min}}$ & Computed $\langle E\rangle\big|_{\beta=100}$ & \% of $E_\text{min}$ & Best found $E_\text{min}$ & \% of $E_\text{min}$ \\
    \hline
    $ 100$ & $-4.77876$ & $-4.74312$ & $99.25$ & $-4.89680$ & $102.47$ \\
    $ 200$ & $-9.65223$ & $-9.35462$ & $96.92$ & $-9.66073$ & $100.09$ \\
    $ 300$ & $-14.5257$ & $-13.9783$ & $96.23$ & $-14.4448$ & $99.44$ \\
    $ 400$ & $-19.3992$ & $-18.5410$ & $95.58$ & $-19.0003$ & $97.94$ \\
    $ 500$ & $-24.2726$ & $-23.2886$ & $95.95$ & $-23.6976$ & $97.63$ \\
    $ 600$ & $-29.1461$ & $-28.0214$ & $96.14$ & $-28.5733$ & $98.03$ \\
    $ 700$ & $-34.0196$ & $-32.7294$ & $96.21$ & $-32.9903$ & $96.97$ \\
    $ 800$ & $-38.8930$ & $-37.4078$ & $96.18$ & $-37.6750$ & $96.87$ \\
    $ 900$ & $-43.7665$ & $-41.7617$ & $95.42$ & $-42.2432$ & $96.52$ \\
    $1000$ & $-48.6400$ & $-46.7495$ & $96.11$ & $-47.1459$ & $96.93$ \\
  \end{tabular}
  \vskip0.5\baselineskip
  \caption{Predicted minimum energy values using~\eqref{eq:best_fit_30} and computed average energy values~\eqref{eq:average_energy} for filaments of length $N$ (at $\beta=100$) and their comparisons.}
  \label{tab:energies_min_100_1000}
\end{table}
The next two columns of the table show the computed values of the average energy $\langle E\rangle$ at $\beta=100$ using the LT algorithm (also seen in Fig.~\ref{fig:results-100-1000_left-side}), which should presumably be close to the minimum values, and their percentage proportions with respect to the predicted minima. Notice that all computed averages are within $5\,\%$ of the predicted minimum values. While computing the average energies, lowest-energy encountered configurations were saved. The energies of the currently lowest-energy configurations, together with their percentage proportions with respect to the predicted minima are shown in the last two columns of the table. Notice that all these results are within $3.5\,\%$ of the predicted values.

\begin{figure}
    \begin{center}
        \includegraphics[width=0.95\textwidth]{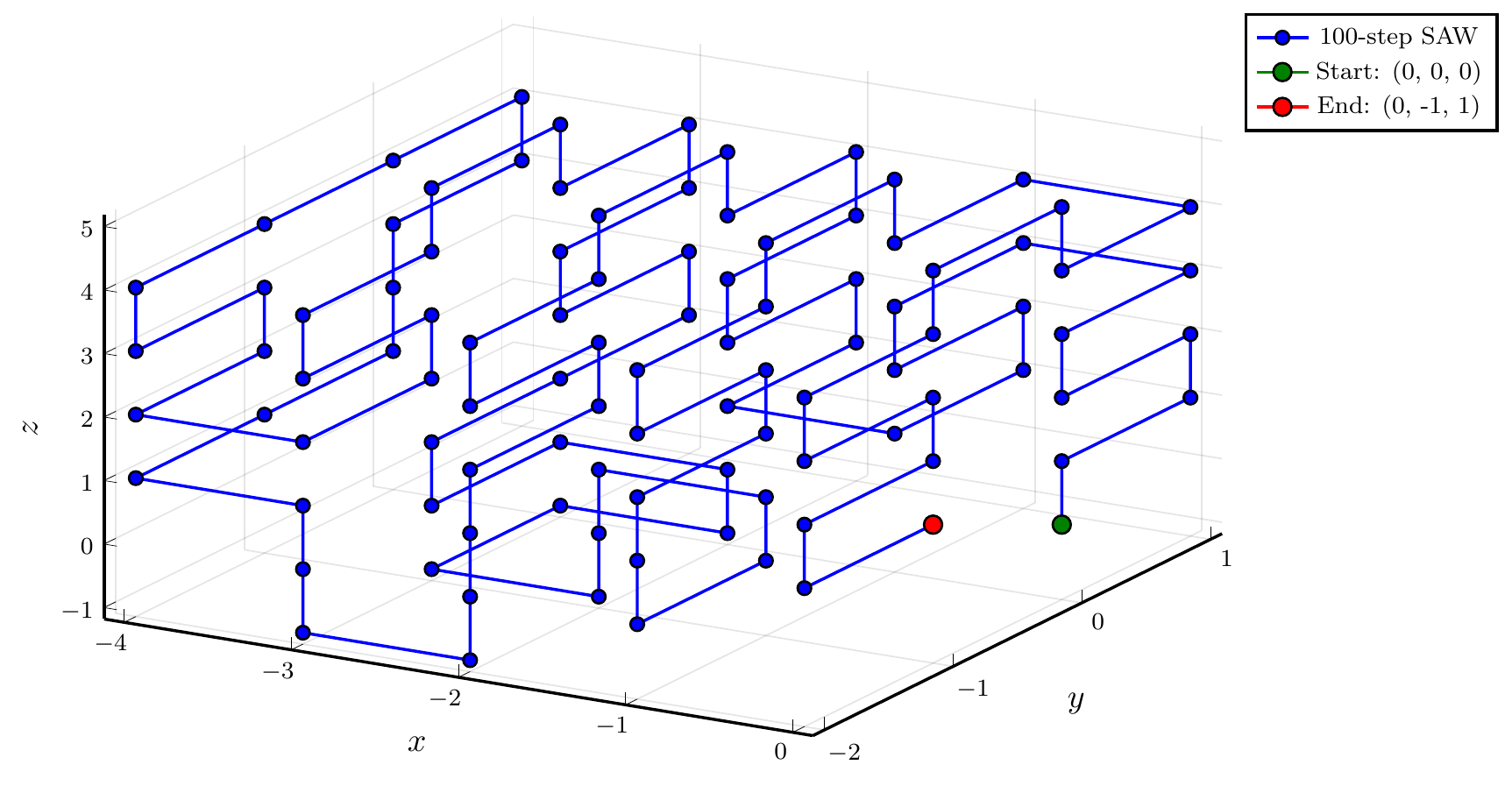}
        \includegraphics[width=0.95\textwidth]{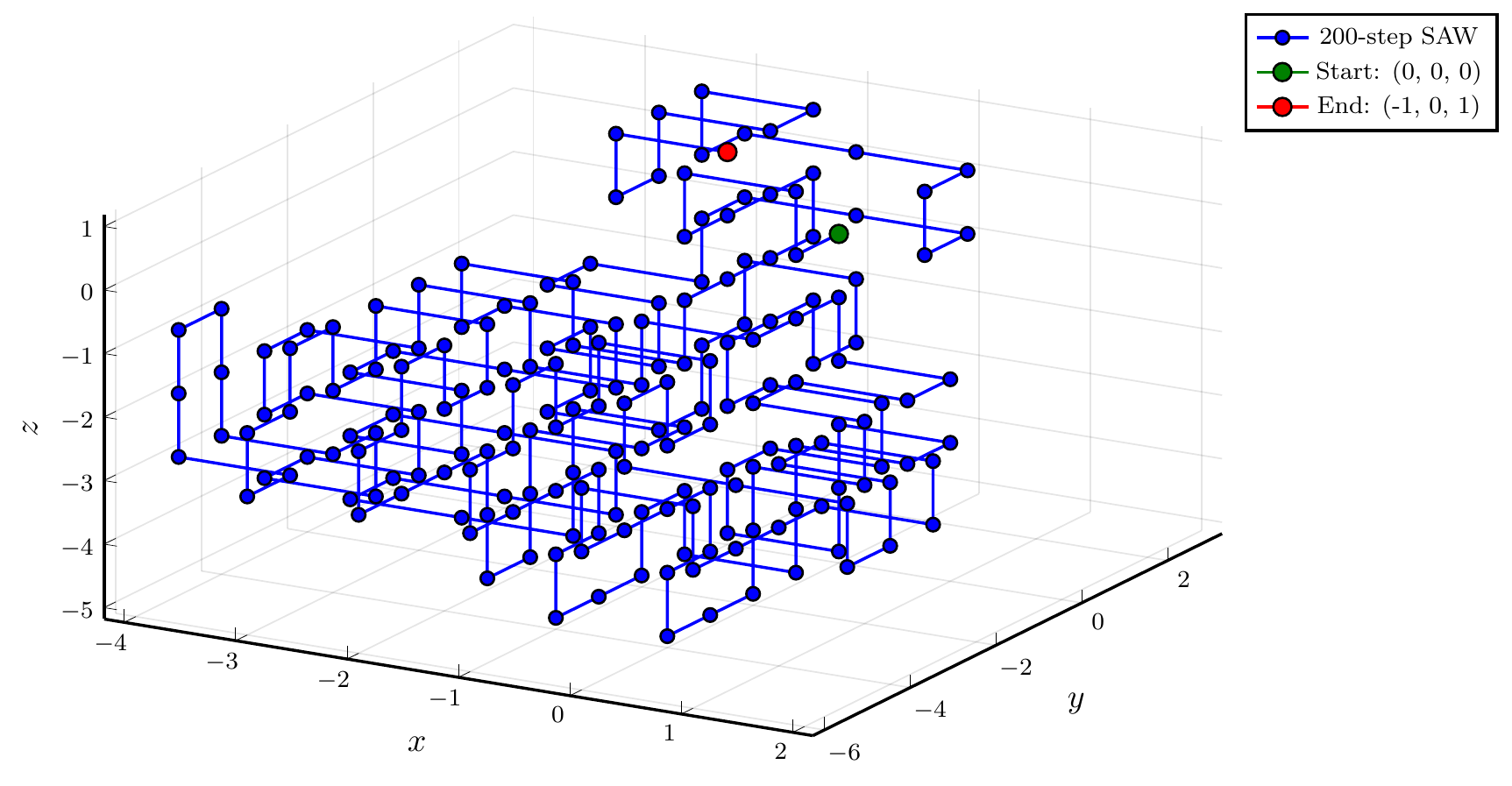}
    \end{center}
    \caption{Lowest energy configurations found by the LT algorithm for $N=100$ and $N=200$. Notice how the endpoints end up the shortest possible distance from each other and also how the SAWs are ``compressed'' into relatively small volumes.}
    \label{fig:saw_min_100-200}
\end{figure}

\begin{figure}
    \begin{center}
        \includegraphics[width=0.95\textwidth]{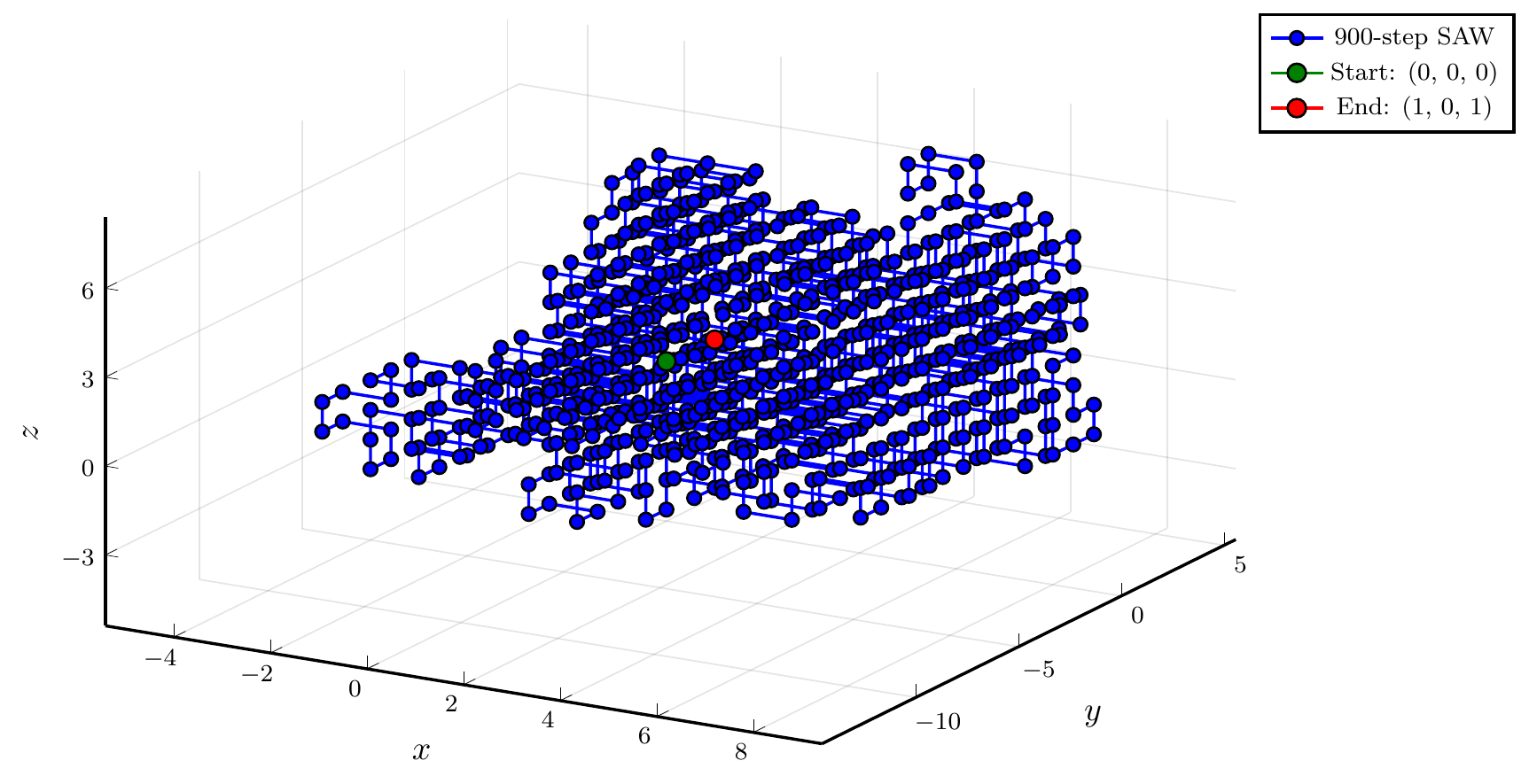}
        \includegraphics[width=0.95\textwidth]{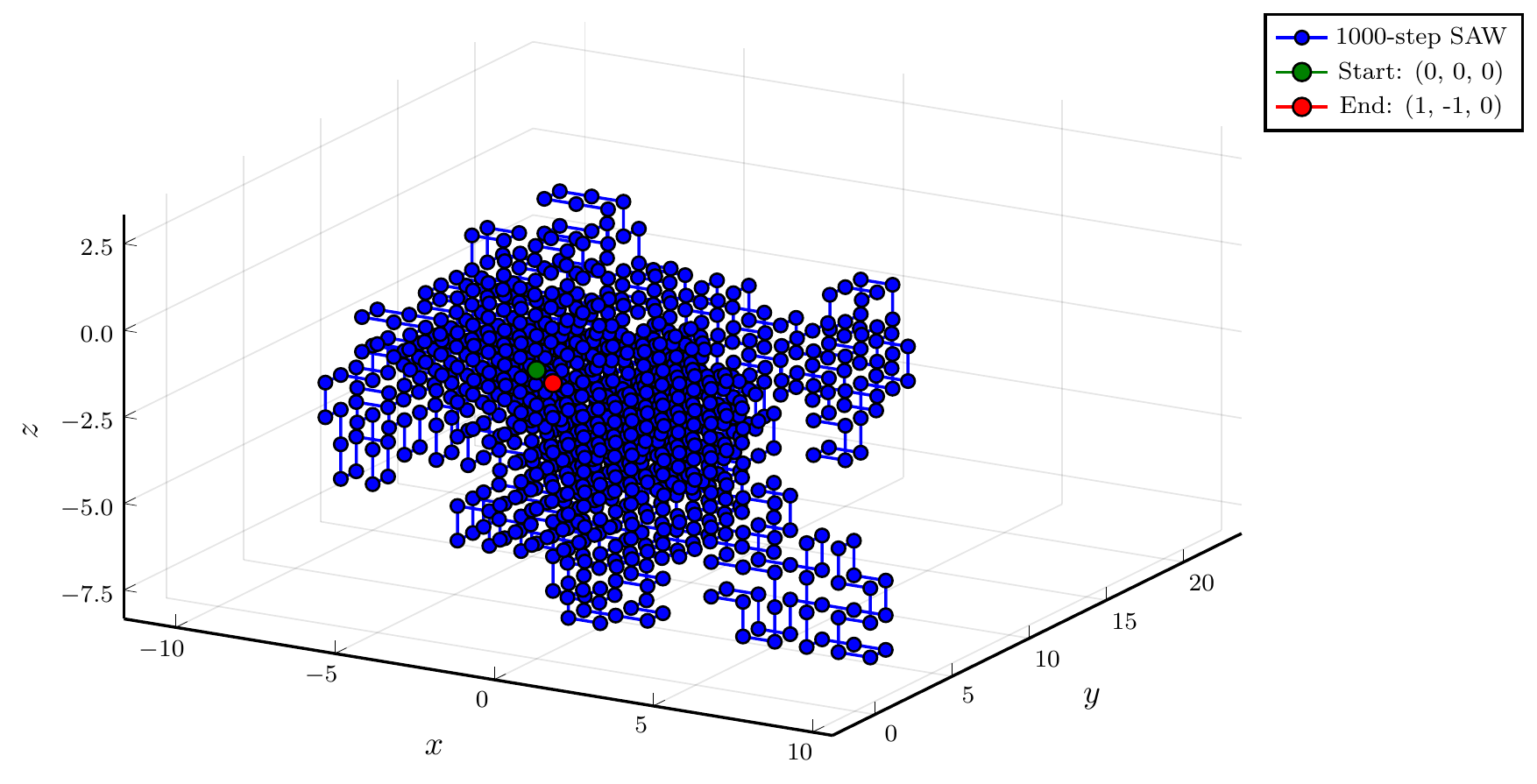}
    \end{center}
    \caption{Same as in Fig.~\ref{fig:saw_min_100-200} but for $N=900$ and $N=1000$.}
    \label{fig:saw_min_900-1000}
\end{figure}

In Figs.~\ref{fig:saw_min_100-200} and \ref{fig:saw_min_900-1000} we show the currently lowest-energy configurations for $N=100$, $200$, $900$, and $1000$ found by the LT algorithm. Notice how these configurations are ``compressed'' into relatively small volumes, suggesting that the true energy-minimizing configurations might also be volume minimizing in some sense. This is also the case for smaller values of $N$ (not shown here). Consequently, in the numerical search for a minimum, one might consider ``outliers,'' such as the part of the filament in the lower-right portion of the plot for $N=1000$ in Fig.~\ref{fig:saw_min_900-1000}, and focus on reconstructing those parts to lie closer to the rest of the filament.

It is also interesting to point out that all the lowest-energy configurations, whether exact or found approximately in our simulations, ended up with their endpoints in the same unit square on the cubic lattice as illustrated in Figs.~\ref{fig:saw_min_100-200} and \ref{fig:saw_min_900-1000}. This observation has not been reported in any of the related works~\cite{chorin88,chorin90,chorin91,chorin,chorinakao91} and it has several consequences. First, any of the quantities computed in~\cite{chorin88,chorin90,chorin91,chorin,chorinakao91} based on the distance between the endpoints of the filaments (denoted by $\mu_{1,N}$, $\mu_{2,N}$, $\mu$, $D_1$, $\tilde{D}$, $\gamma$, etc.~in the references) need to be computed in a different way, and therefore we will not attempt to reproduce those results here. Second, it is likely that this observation is a general attribute of energy-minimizing configurations and worth proving analytically. We have not succeeded in this effort. Third, the numerical search for energy-minimizing configurations may perhaps be done more efficiently using this observation. It limits the set of possible filament configurations that need to be considered, but it also offers a different point of view: consider such filaments as closed (also known as self-avoiding polygons) and then remove one or two segments, depending on the parity of $N$. We used the first idea to slightly expand the list of values in Table~\ref{tab:energies_min} to obtain the results for $19\le N\le21$, which were then matched by the results of the LT algorithm.

\section{Conclusions}
\label{sec:conclusions}
In this work we proposed a new algorithm, referred to here as the {\it localized transformations} (LT) algorithm, for the computation of statistical equilibrium quantities on a cubic lattice when both an energy and a statistical temperature are involved. We demonstrated that the pivot algorithm used in situations such as protein folding works well for a small range of temperatures near the polymeric case, but it fails in other situations. Specifically, we have observed that non-straight configurations at large (in absolute value) negative values of $\beta$ (for which the Boltzmann probability distribution~\eqref{eq:boltzmann_distribution} strongly favors high-energy, straight configurations) are extremely unlikely to straighten out in the MCMC simulation since intermediate steps with non-local transformations are required that greatly affect the energy. Similarly, when starting with a straight configuration (that can be viewed as corresponding to a large, in absolute value, negative $\beta$), the same need for an intermediate transformation with a large energy change affects the results of the MCMC simulations at smaller (in absolute value) values of $\beta$.

At the other end of the inverse temperature spectrum, when the values of $\beta$ are large positive, the Boltzmann probability distribution strongly favors lowest-energy configurations, likely very much folded up and compressed into a small volume. While the maximum-energy configurations are obviously the straight configurations with their energies easily computed for any $N$, the minimum-energy configurations are not known to us, nor are the actual minimum energies. However, based on the results shown in section~\ref{sec:min_energy}, the MCMC simulation based on the pivot algorithm struggles to get anywhere close to the projected minimum values, thus begging for a different approach to the problem. Part of the problem with the pivot algorithm again is the need for intermediate configurations with unfavorable values of the energy; another problem, still likely present in our approach, is that the minimum-energy configurations appear to require a particular structure that may be hard to get to from another folded configuration even when using simple local transformations.

The proposed LT algorithm seems to perform well for all possible temperature values. The difficulties with the pivot algorithm outlined in the previous paragraphs, when energy and the Boltzmann probability distribution are involved, suggest that the pivot algorithm is too rigid in the non-polymeric cases ($\beta$ away from $0$, or $T$ away from $\infty$), and that different, more local, transformations need to be considered in order to better approximate average energies when $\beta\ne0$. To this end, we proposed and utilized two types of transformations: one designed specifically
to help with straightening out non-straight configurations in order to increase their energy at negative values of $\beta$, and one designed to help with compressing configurations in order to lower their energy at positive values of $\beta$. The latter was motivated by previous work for generating self-avoiding walks with fixed endpoints~\cite{madrasorlitskyshepp89}.

Having reliably approximated the values of equilibrium energy, we also proposed an efficient way to compute equilibrium entropy for all temperature values. This algorithm mimics Euler's method in that it approximates the solution to $\partial S/\partial\langle E\rangle=k_B\beta$ instead of computing the entropy directly, and it requires the knowledge of $\langle E\rangle$ for a range of values of $\beta$, as well as an initial value of the entropy $S$ for one value of $\beta$.

Finally, we applied the algorithms in the context of suction or supercritical vortices in a tornadic flow, which are approximated by vortex filaments on a cubic lattice. We confirmed that the supercritical vortices (smooth, ``straight'' vortices) have the highest energy and correspond to negative temperatures in this model. The lowest-energy configurations are folded up and compressed to a great extent. From the point of view of the flow of energy, the negative-temperature, high-energy suction vortices are expected to lose (some of) their energy to the surrounding flow. The results also support A.~Chorin's findings that in the context of supercritical vortices in a tornadic flow, when such high-energy vortices stretch, they need to fold. Specifically, as can be seen in Fig.~\ref{fig:results-100-1000}, with a fixed energy $\langle E\rangle$ and at negative temperatures, longer vortices (larger $N$) are farther away from their maximum energies than shorter vortices, and thus farther away from straight configurations, which corresponds to more folds present along such vortices.

\bibliography{saw}
\bibliographystyle{abbrv}

\end{document}